 \documentclass[sigconf]{acmart}  

\usepackage[ruled,vlined,linesnumbered]{algorithm2e}
\usepackage{amsmath}
\usepackage{color}
\usepackage{graphicx}
\usepackage{rotating}
\usepackage{pifont}% http://ctan.org/pkg/pifont
\definecolor{munsell}{rgb}{0.13, 0.55, 0.13}
\newcommand{\cmark}{\color{munsell}\ding{51}}%
\newcommand{\xmark}{\color{red}\ding{55}}%

\DeclareMathOperator*{\argmin}{arg\,min}
\newcommand\add[1]{#1}
\newenvironment{ADD}{\par}{\par}

\usepackage{marginnote}

\AtBeginDocument{%
  \providecommand\BibTeX{{%
    \normalfont B\kern-0.5em{\scshape i\kern-0.25em b}\kern-0.8em\TeX}}}

\newcommand{\name}[1]{\def\papername{#1}}
\name{Graph4GUI}
\acmSubmissionID{}

\copyrightyear{2024}
\acmYear{2024}
\setcopyright{rightsretained}
\acmConference[CHI '24]{Proceedings of the CHI Conference on Human Factors in Computing Systems}{May 11--16, 2024}{Honolulu, HI, USA}
\acmBooktitle{Proceedings of the CHI Conference on Human Factors in Computing Systems (CHI '24), May 11--16, 2024, Honolulu, HI, USA}
\acmDOI{10.1145/3613904.3642822}
\acmISBN{979-8-4007-0330-0/24/05}

\begin{document}


\title{Graph4GUI: Graph Neural Networks for Representing Graphical User Interfaces}

\author{Yue Jiang}
\email{yue.jiang@aalto.fi}
\affiliation{
    \institution{Aalto University} 
    \country{Finland}
}

\author{Changkong Zhou}
\email{changkong.zhou@aalto.fi}
\affiliation{
    \institution{Aalto University} 
    \country{Finland}
}

\author{Vikas Garg}
\email{vgarg@csail.mit.edu}
\authornote{Co-last authors, equal contribution.}
\affiliation{
    \institution{YaiYai Ltd and Aalto University} 
    \country{Finland}
}

\author{Antti Oulasvirta}
\email{antti.oulasvirta@aalto.fi}
\authornotemark[1]
\affiliation{
    \institution{Aalto University} 
    \country{Finland}
}

% \affiliation{%

% \affiliation{%

% \affiliation{%

% \affiliation{%

% \affiliation{%



%%
%% By default, the full list of authors will be used in the e props
\renewcommand{\shortauthors}{Jiang et al.}

\begin{abstract}          
Present-day graphical user interfaces (GUIs) exhibit diverse arrangements of text, graphics, and interactive elements such as buttons and menus, but representations of GUIs have not kept up. They do not encapsulate both semantic and visuo-spatial relationships among elements. %\color{red} 
To seize machine learning's potential for GUIs more efficiently, \papername~ exploits graph neural networks to capture individual elements' properties and their semantic—visuo-spatial constraints in a layout. The learned representation demonstrated its effectiveness in multiple tasks, especially generating designs in a challenging GUI autocompletion task, which involved predicting the positions of remaining unplaced elements in a partially completed GUI. The new model's suggestions showed alignment and visual appeal superior to the baseline method and received higher subjective ratings for preference. 
Furthermore, we demonstrate the practical benefits and efficiency advantages designers perceive when utilizing our model as an autocompletion plug-in.

\end{abstract}

\newcommand{\loss}{\mathcal{L}}
\newcommand{\image}{\mathcal{I}}
\newcommand{\encoder}{\mathcal{E}}
\newcommand{\decoder}{\mathcal{D}}
\newcommand{\normal}{\mathcal{N}}

\begin{CCSXML}
<ccs2012>
   <concept>
       <concept_id>10003120.10003121.10003129</concept_id>
       <concept_desc>Human-centered computing~Interactive systems and tools</concept_desc>
       <concept_significance>500</concept_significance>
       </concept>
   <concept>
       <concept_id>10003120.10003121.10003128</concept_id>
       <concept_desc>Human-centered computing~Interaction techniques</concept_desc>
       <concept_significance>500</concept_significance>
       </concept>
 </ccs2012>
\end{CCSXML}

\ccsdesc[500]{Human-centered computing~Interactive systems and tools}
\ccsdesc[500]{Human-centered computing~Interaction techniques}

\keywords{Graphical User Interface, User Interface Representation, Constraint-based Layout, Graph Neural Networks}

\begin{teaserfigure}
  \def\w{\linewidth}
  \centering
 \includegraphics[width=\w, trim=0 0 0 0]{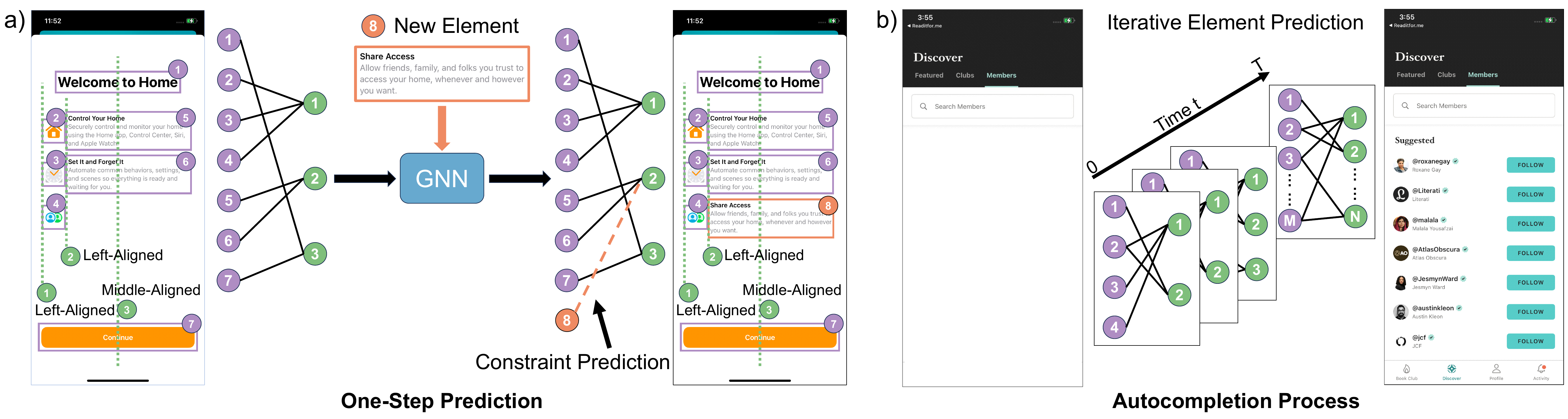}
\caption{
Graph4GUI is a graph-based GUI representation that captures the connections between GUI element properties and constraints.
Such representation can capture the visual--spatial--semantic structure of a GUI such that it could be effectively employed in
computational design. a) To represent the GUIs, bipartite graphs comprising element nodes (colored purple) convey the GUI
elements' properties and constraint nodes (colored green) that can be integrated into graph neural networks. Such representation can
serve various downstream tasks, such as predicting constraints (dotted orange edge) for an unplaced element (colored orange). b) By iteratively predicting
the sizes and locations of yet-unplaced elements, we can support designers by autocompleting partially completed GUI designs.
%We propose a graph-based GUI representation to establish the connections between GUI element properties and constraints. Such representation can capture the visual–spatial–semantic structure of a GUI such that it could be effectively employed in computational design. a) We represent GUIs as bipartite graphs, comprising element nodes (colored purple) that convey the GUI element properties and constraint nodes (colored green) that can be integrated into graph neural networks. Such representation can be used for various downstream tasks, such as predicting constraints (blue edge) for an unplaced element. b) By iteratively predicting the sizes and locations of yet-unplaced elements, we can support designers by autocompleting partially completed designs.
}
  \Description{We show that Graph4GUI is a graph-based GUI representation connecting GUI element properties and constraints. a) 
We use bipartite graphs comprising element nodes to convey the GUI
elements' properties and constraint nodes (colored green) to represent the GUIs. It can predict the edge between element nodes and constraint nodes for an unplaced element. b) By iteratively predicting
the sizes and locations of yet-unplaced elements, we can support designers by autocompleting partially completed GUI designs.}
  \label{fig:teaser}
\end{teaserfigure}

\maketitle

\begin{figure*}[t]
  \def\w{\linewidth}
  \centering
 \includegraphics[width=\w]{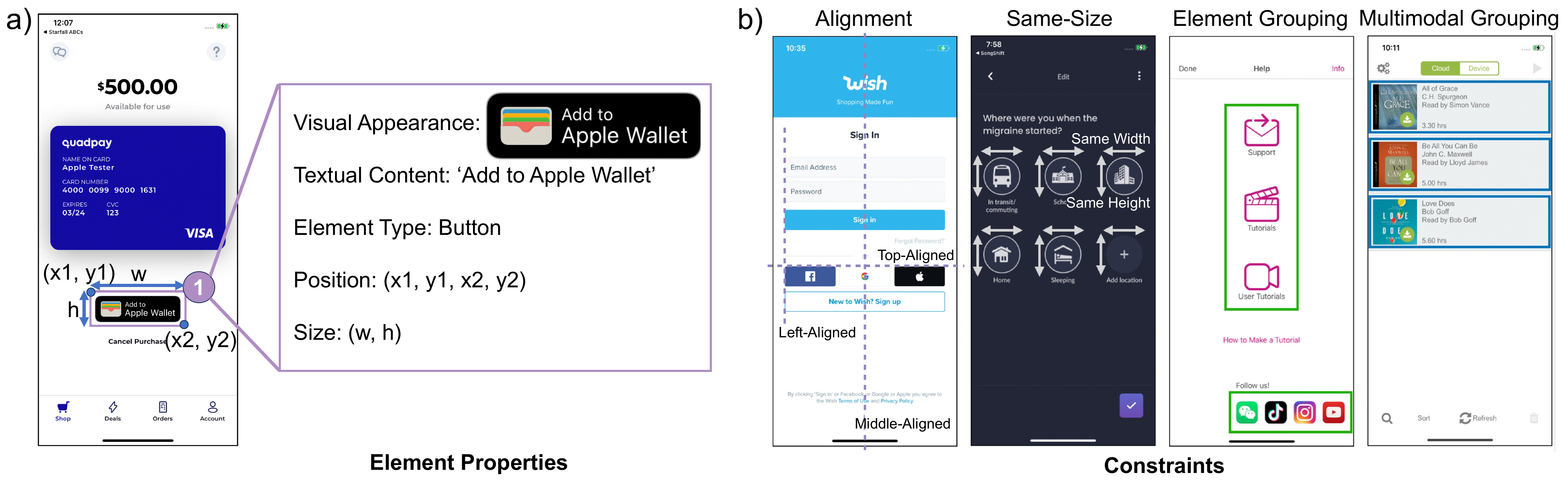}
\caption{\add{a) \papername~ represents each GUI element through a separate node with properties. GUI element nodes convey the element properties, including visual appearance, textual content, element type, position, and size. b) Constraint nodes express four types of constraints: alignment, same-size, element grouping, and multimodal grouping constraints.}}
\Description{a) In the element node, we consider element properties, including visual appearance, textual content, element type, position, and size. b) Constraint nodes express four types of constraints: alignment, same-size, element grouping, and multimodal grouping constraints.}
  \label{fig:element_and_constraints}
\end{figure*} 

\section{Introduction}
 
Modern graphical user interfaces (GUIs) are replete with diverse elements like text, graphics, buttons, checkboxes, sliders, and icons, arranged in various ways.
GUIs deploy visual, spatial, and textual cues to guide users in their design.
For example, colors convey grouping and attention, while visual cues like proximity or shared visual areas signal element associations~\cite{Brumby2015VisualGI}.
Elements are ordered and grouped based on grid lines; for instance, lists are often left-aligned \cite{miniukovich2015computation}.
In addition, textual elements, such as headers, labels, and annotations, are needed to communicate the ``semantics'' of the various shapes and images.
Despite architectural commonalities, each GUI genre and application has its unique conventions.
The question of how to represent a GUI's visual--spatial--semantic structure such that it could be effectively conveyed in computational design remains open~\cite{jiang2022computational, jiang2023future, jiang2024computational, jiang2024computational2}.

Prior methods for representing GUIs and their constituent elements fall short of capturing these integrative aspects. 
Some work has focused exclusively on textual content in GUIs, but neglected the visual aspects of design and the variety of GUI elements~\cite{li2017sugilite, li2020multi}.
In contrast, other approaches emphasize visual appearance and GUI element types but overlook the content of the elements~\cite{deka2016erica, learning2022ang, learning2018liu}. This results in similar treatment for GUIs sharing structural and visual similarities but differing in content. 
\add{Layout constraints represent the layout relationships between GUI elements, such as alignment, same-size, and grouping. Most existing methods, employing Convolutional Neural Networks (CNNs) to learn GUI images, face challenges because they have to learn layout constraints from pixels. This makes training a model to predict constraints a challenge.}

To address this gap, we propose a novel graph-based GUI representation \papername~ that integrates GUI elements with layout constraints (Figure~\ref{fig:teaser}a). We formulate a bipartite graph to express GUI elements and their relationships via two kinds of nodes: element and constraint nodes. \add{As shown in~\autoref{fig:element_and_constraints}, each GUI element node represents element properties, including visual appearance, textual content, element type, position, and size. Constraint nodes include four types of constraints: alignment, same-size, element grouping, and multimodal grouping constraints.}
We then employ Graph neural networks (GNNs) to learn a domain-specific representation from the graph-structured data. 

\add{
Compared to other GUI representations~\cite{li2017sugilite, li2020multi, deka2016erica, learning2022ang, learning2018liu, dayama2020grids, screen2vec2021li}, the design of our GNN aims to balance two representation learning goals.
On the one hand, we want to maximally exploit domain knowledge, particularly using stable, universal GUI design characteristics without learning from scratch. On the other, we want to capture contingent design tendencies -- such as color palettes and fonts -- without manual specification. If successful, this would make it possible to learn a useful representation with fewer samples.
To this end, our approach is to represent relatively universal principles of layouts as constraints in a graph, thus reducing the need to learn them from scratch. At the same time, genre-specific tendencies are learned by applying GNNs to capture the design features unique to GUIs. In contrast, traditional structured representations in computational design, such as DOM trees, are not designed for learning but for specifying GUIs. While they represent the view hierarchy of a layout, they do not lend themselves to many machine learning methods. For example, there is no natural way to represent the concept of grid alignment with DOMs, as this information is split into the leaf nodes of the tree. To compute whether two elements are aligned, the whole tree needs to be parsed. In contrast, our method directly embeds connections between GUI elements and their constraints in the graph.
}

\begin{ADD}
\def\arraystretch{1}% 
\begin{table*}[t!]
% \marginparsep=30pt
% \marginnote{\color{violet}
% 1AC, 2AC, R2, R3: Added Table 1 to clarify that our method consider more aspects of the GUI compared to other methods.\\~\\}
\scalebox{1}{\setlength\tabcolsep{3pt}
\begin{tabular}{lcccccc}
\hline
\add{\textbf{Approach}} &  \add{\textbf{Textual Content}} & \add{\textbf{Visual Appearance}} & \add{\textbf{Element Type}} & \add{\textbf{View Hierarchy}}  & \add{\textbf{Layout Constraints}}   \\
\midrule
\add{SUGILITE~\cite{li2017sugilite}} & \cmark & \xmark & \xmark & \cmark  & \xmark   \\
\add{SOVITE~\cite{li2020multi}} & \cmark   & \xmark & \xmark & \cmark & \xmark  \\
\add{ERICA~\cite{deka2016erica}} & \xmark & \cmark  & \cmark   & \cmark  & \xmark  \\
\add{HAMP~\cite{learning2022ang}} & \xmark & \cmark  & \cmark  & \xmark & \xmark  \\
\add{Liu et al.~\cite{deka2016erica}} & \xmark & \cmark   & \cmark  & \cmark  & \xmark \\
\add{Screen2Vec~\cite{screen2vec2021li}} & \cmark  & \xmark & \cmark  & \cmark  & \xmark \\
\add{Li et al.~\cite{li2020auto}} & \xmark & \xmark & \cmark  & \cmark   & \xmark \\
\add{Br{\"u}ckner et al.~\cite{bruckner2022learning}}  & \xmark & \xmark & \cmark  & \xmark & \xmark  \\
\add{GRIDS~\cite{dayama2020grids}}  & \xmark & \xmark & \cmark  & \xmark & \cmark   \\
\midrule
\add{Ours}  & \cmark  & \cmark  & \cmark  & \xmark & \cmark  \\
\bottomrule
\end{tabular}
}
\caption{\add{A comparison of existing approaches, marked based on if they \textit{explicitly represent}: textual content, visual appearance, GUI element type, view hierarchy, and layout constraints. The view hierarchy, akin to a DOM tree, begins with a root view and organizes all its descendants in a tree structure. Layout constraints denote relationships like alignment and grouping among GUI elements. ``\cmark\color{black}'' indicates that the model can capture the factor, while ``\xmark\color{black}'' indicates that it does not.}}
\Description{This table shows whether existing approaches explicitly represent textual content, visual appearance, GUI element type, view hierarchy, and layout constraints.}
\label{tbl:related_work_comparison}
\end{table*}
\end{ADD}

\add{To examine the effectiveness of our graph-based representation, we applied it to different applications: GUI autocompletion, GUI topic classification, and GUI retrieval.
Our primary emphasis lies in autocompletion due to its complexity.
This autocompletion problem is challenging, not only because exploring potential GUI element combinations is computationally costly but also because good solutions must consider visual, spatial, and semantic constraints among the to-be-placed elements and those already present.} 
\add{We present a method for iteratively recommending the position and size of unplaced GUI elements, as illustrated in Figure~\ref{fig:teaser}b. To augment the model's usability for designers, we introduce alternative element suggestion options, including the recommendation of element groups and simultaneous suggestions for all elements.}
For evaluation, we conducted Study 1, comparing it with GRIDS~\cite{dayama2020grids}, an approach for autocompletion using integer programming. Our model produced suggestions with superior alignment and visual appeal compared to the baseline, consistent with participants' preferences. In Study 2, we integrated our model into a plug-in for a design tool. It allows GUI designers to utilize autocompletion capabilities in real time while maintaining full control over the design process within the interactive design tool. We interviewed six GUI designers to study our tool's practical benefits and efficiency advantages.

Our work makes the following contributions:
\begin{enumerate}
\item A novel graph representation for GUIs, \papername, which incorporates GUI element properties such as textual content, visual appearance, and element types, along with their relationships and constraints.
\item A graph neural network method for learning the graph representation of the GUI to optimize the GUI element dimensions and positions.
\item An autocompletion framework that serves to demonstrate the graph representation facilitating interactive GUI design. The framework's effectiveness was evaluated through a comparison study and a designer study.
\item \add{Applying the graph representation to other applications, including GUI topic classification and GUI retrieval.}
\end{enumerate}

\section{Related Work}

This section focuses on the limitations of preexisting representations of GUIs, the GUI-related applications of graph neural networks, and constraint-based approaches to layout generation.

\subsection{Representations of GUIs}

Existing GUI representations often prioritize specific properties while neglecting others. \add{Table \ref{tbl:related_work_comparison} provides a comparison of existing approaches, marked based on whether they explicitly represent textual content, visual appearance, GUI element type, view hierarchy, and layout constraints.
Some representations focus on textual content, ignoring visual appearance and GUI element types~\cite{li2017sugilite, li2020multi}.
In contrast, alternative methods prioritize visual appearance and the types of GUI elements~\cite{deka2016erica, learning2022ang, learning2018liu}, but often overlook the importance of textual content. This can lead to similar treatment of structurally and visually similar GUIs that differ significantly in textual content.}
Screen2Vec~\cite{screen2vec2021li} addresses this by generating GUI representations incorporating textual content, element types, and screen hierarchy. Our method, Graph4GUI, extends this consideration by incorporating constraints and interrelationships between GUI elements. ILuvUI~\cite{jiang2023iluvui} proposed a vision language model to create a language representation of GUI.
\add{A relevant method, GRIDS~\cite{dayama2020grids}, is an integer programming method optimizing grid layouts using layout constraints. 
 We also consider layout constraints since they are important in developing a well-structured GUI design.
 }
With our approach, Graph4GUI, we propose a solution that considers not only textual content, visual appearance, and GUI element types but also the constraints and interrelationships between GUI elements. 

\subsection{Graph Neural Networks on GUIs}

Graph neural networks~\cite{gori2005new, scarselli2009graph, garg20GNNs, xu2018how} are state-of-the-art models for encoding graph-structured data. Whereas CNNs rely on convolution over spatial neighborhoods and enjoy widespread application to encode GUI images, GNNs aggregate information from neighborhoods defined by an input graph that are not restricted to the spatial domain. This gives them the potential to exploit information about the GUIs beyond pixel level. 
\add{Li et al. applied GNNs to denoise an existing user-interface dataset~\cite{li2022learning}, and performed GUI autocompletion from the GUI layout hierarchy but failed to generate visually realistic GUI results~\cite{li2020auto}. Br{\"u}ckner et al.~\cite{bruckner2022learning} looked into constructing a graph using GUI elements' relative positioning to predict elements; however, it proved challenging to learn the layout structure from only relative positions.
HAMP~\cite{ang2022learning}, introduced a graph representation with nodes for app descriptions, GUI screens, GUI classes, and element images to perform GUI tasks. Still, such detailed metadata cannot be extracted from GUI screenshots without extensive manual annotations.
In contrast, our application of GNNs is geared toward modeling the layout graph of GUI elements, thereby enabling us to capture both the topological intricacies of the GUI layout and the properties of individual GUI elements.}

\subsection{Constraint-based Layout Generation}
 
Constraint-based layout models have been widely used in GUI layouts~\cite{sahami2013insights, zeidler2017automatic, marcotte2011responsive, badros2001cassowary, bill1992bricklayer, borning1997solving, hosobe2000scalable, lutteroth2008domain, sadun2013ios, weber2010reduction, zeidler2017tiling, karsenty1993inferring, scoditti2009new, zeidler2012auckland, borning1968constraint, szekely1988user} and document layouts~\cite{hurst2003cobweb, hosobe2005solving, borning2000constraint, laine2021responsive}. \add{Early methods like Peridot~\cite{myers1986creating, myers1990creating} and Lapidary~\cite{zanden1991lapidary} proposed programming by demonstration, automatically generate constraints for user interfaces based on designer interactions.} These models offer greater flexibility for layout generation than simple layout models such as group, grid, table, and grid-bag layouts~\cite{myers2000past, myers1995user, myers97theamulet}. 
Prior work proposed constraint-based layout generation~\cite{weld2003automatically, fogarty2003gadget}.
For instance, SUPPLE~\cite{Gajos2008decision, gajos2004supple, gajos2008improving} presented constraints for alternative widgets and groupings, and
ORCLayout~\cite{jiang2019orclayout, jiang2020orcsolver, jiang2020reverseorc} introduced OR-constraint as a mixture of hard and soft constraints to unify flow-based and constraint-based layouts. 
\add{Constraints have functioned also to enable layout personalization~\cite{Gajos2005preference}, maintaining consistency~\cite{gajos2005cross}, giving layout-alternative suggestions based on user-defined constraints~\cite{swearngin2020scout, bielik2018robust}, generating layout alternatives from templates or modifiable suggestions~\cite{Jacobs2003document, sinha2015responsive, zanden1990automatic}, and allowing both author and viewer to specify the layout~\cite{borning2000constraint}.
Finally, recent work has explored applying deep-learning approaches to automatic layout generation, eliminating the need for manually defined constraints~\cite{zheng2019content, lee2019neural}.}
However, none of these methods predict constraints for GUI elements. 
Incorporating GUI element relationships as constraints enables our model to predict them within the network. This enhances the network's ability to establish connections and deepen its comprehension of both element properties and constraints.

\section{GUI Layout Problem}
\label{objective}

Designing a GUI involves carefully selecting elements and organizing them into a structure that is usable and aesthetically appealing. This brings with it a large number of both element-specific and layout-related decisions and is typically iterative in nature~\cite{galitz2007essential}. 
Our objective is to provide a more comprehensive characterization of GUIs compared to existing GUI representations by factoring in visual, spatial, and semantic features. We formulate the problem by partitioning it into elements and layouts while also defining the GUI design problem as an optimization process. 

\subsection{Element Properties}

The first major consideration is \textit{visual appearance}, a broad notion encompassing such properties as color combinations, geometric shapes, and GUI styles. For example, employing tranquil hues such as blues and greens may encourage a calming interface, requiring a more spacious and streamlined layout. In contrast, energetic shades (reds, yellows, etc.) might necessitate a more condensed and high-energy design. Likewise, the arrangement of the various shapes plays a crucial role in the overall visual appeal of the layout.
Equally essential is the \textit{textual content} -- encompassing all forms of text content visible in the user interface. Labels can play a crucial role from the design perspective. 
The alignment and arrangement of labels that contain long paragraphs require larger spaces, to avoid clutter and overlapping. Conversely, elements containing brief text strings or bullet-point-style items may allow for compaction, thereby leaving room for other pertinent features. The format of the content is also critical; condensed or expanded layouts especially require careful consideration of element spacing, alignment, and the overall design arrangement.
Finally, each \textit{element type}, such as button, checkbox, or text field, has inherent functionality and objectives. For example, buttons need to be easily reachable if user interaction is to be effective, whereas checkboxes might not require such prominent positioning on account of their less frequent use. Text fields' usual dominance as the focal point is due to their role in data entry, which necessitates adequate design space.

\subsection{Layout Constraints}

Layout-level properties can be approached as constraints. Imposing constraints can help maintain consistency across GUIs and aid users in understanding them. Among commonplace GUI constraints are keeping similar elements the same size, aligning elements along a shared grid, and grouping related elements together.
The \textit{alignment} constraint, for instance, enforces uniform positioning and visual consistency by arranging elements along a shared axis, thus maintaining a structure within the layout~\cite{miniukovich2015computation}. In addition to establishing relationships among elements, alignment strengthens the synergy between the elements and the broader layout.
The \textit{same-size} constraint is equally essential in that it guarantees maintaining appropriate sizes across GUI elements. This enhances visual harmony by making sure of consistent sizing among related elements.
\textit{Element grouping} is important for enhancing the layout's organization and logical structure. This is achieved by bringing together related elements with similar functions. The strategy promotes user-friendly navigation.
Finally, \textit{multimodal grouping} constraints lend a structured feel to varied elements, with coherent organization across distinct types. This allows for a harmonious combination of text, images, and other GUI components while remaining respectful of the layout's coherence and uniformity principles.

\subsection{Formulation of GUI Layout Problem}
\label{sec:objective_function}

We can now define the GUI layout problem as an optimization problem. 
With this formulation, we decide on the positions and sizes of elements in a GUI, denoted as ${e}_i = ({x}_i, {y}_i, {w}_i, {h}_i)$, where the coordinates $(x_i, y_i)$ represent the top-left corner of the $i$-th element and $(w_i, h_i)$ represents its width and height. Here, we focus on a setting where all the elements are in rectangular bounding boxes.

%\color{blue} 
%[For the following part, we first need to introduce the element loss and constraint loss terms. Then only  we should proceed to defining $\mathcal{L}$. The optimization goal should only be stated then. In general, always introduce notations/definitions/descriptions before using them. Also, don't abbreviate $\mathcal{L}(\hat{e}_1, \hat{e}_2, ..., \hat{e}_N, \mathbf{F})$ to $\mathcal{L}$.]\\
%\color{black}
%The optimization goal is:

% \begin{equation}
% \label{eq:optimization}
% (\hat{e}_1^*, \hat{e}_2^*, \hat{e}_N^*) = \argmin_{\{\hat{e}_1, \hat{e}_2, \hat{e}_N\}}  \mathcal{L}.
% \end{equation}

%\color{blue} [$\mathcal{L}$ here is a function of the variables $\hat{e}_1, \hat{e}_2, \ldots, \hat{e}_n$ that we should optimize over and not $e_1, e_2, \ldots, e_n$ that you later define to be ground truth. Also, replace $e_1^*, e_2^*, e_N^*$ etc. by $\hat{e}_1^*, \hat{e}_2^*, \hat{e}_N^*$, and use $({e}_1^*, {e}_2^*, ..., {e}_N^*)$ instead of $\{{e}_1^*, {e}_2^*, ..., {e}_N^*\}$. Please be careful/precise with use of notation. ]  \color{black}

% The objective function $\mathcal{L}$ is used to optimize the predicted positions and sizes of GUI elements $\{{e}_1^*, {e}_2^*, ..., {e}_N^*\}$.

%\subsubsection{Objective Function Formulation for GUI Optimization}

We define two objective terms, the element loss term ($\mathcal{L}_{\mathrm{ele}}$) and the constraint loss term ($\mathcal{L}_{\mathrm{cons}}$). The first of these encapsulates the properties of the GUI elements, such as visual appearance, texture content, and element type. The constraint loss term covers the layout constraints that guide the GUI design toward an optimal arrangement.
The objective function we defined above becomes 

\begin{equation}
\label{eq:objective_function}
\begin{split}
\mathcal{L}(\hat{e}_1, \hat{e}_2, ..., \hat{e}_N, \mathbf{F}; \lambda, \eta) = \mathcal{L}_{\mathrm{ele}}({e}_1, {e}_2, ..., {e}_N, {\hat{e}_1, \hat{e}_2, ..., \hat{e}_N}; \eta)\\ +\ \lambda ~\mathcal{L}_{\mathrm{cons}}(\hat{e}_1, \hat{e}_2, ..., \hat{e}_N, \mathbf{F}),
\end{split}
\end{equation}

% \color{blue} [We can scale $\mathcal{L}_{\mathrm{cons}}$ by $\lambda > 0$ to have a more general form that does not have to treat the two loss terms as equally important:
% $$  \mathcal{L}(\hat{e}_1, \hat{e}_2, ..., \hat{e}_N, \mathbf{F}; \lambda, \eta) = \mathcal{L}_{\mathrm{ele}}({e}_1, {e}_2, ..., {e}_N, {\hat{e}_1, \hat{e}_2, ..., \hat{e}_N}; \eta) + \lambda \mathcal{L}_{\mathrm{cons}}(\hat{e}_1, \hat{e}_2, ..., \hat{e}_N, \mathbf{F})$$

% We can then say later in the experiments section that we set $\lambda$ to 1.
%  ] \color{black}

where $\mathbf{F} = \{f_1, f_2, ..., f_N\}$ pertains to the set of properties for the $N$ GUI elements, including the visual appearance, textual content, and element type. 
The predicted size and position of each GUI element are denoted as $\hat{e}_i = (\hat{x}_i, \hat{y}_i, \hat{w}_i, \hat{h}_i)$, while the ground-truth sizes and positions are represented by $e_i = (x_i, y_i, w_i, h_i)$. $\lambda > 0$ is the weight of the constraint loss. $\eta>0$ is the weight of the boundary loss as a part of the element loss term described below.

The element loss term ($\mathcal{L}_{\mathrm{ele}}$) refers to the discrepancy between the predicted and actual values for both positions and sizes of the GUI elements, with a penalty imposed if the predicted elements are outside the interface area:

\begin{equation}
\label{eq:element_loss_term}
\begin{split}
\mathcal{L}_{\mathrm{ele}}({e}_1, {e}_2, ..., {e}_N, {\hat{e}_1, \hat{e}_2, ..., \hat{e}_N}; \eta)& \\ = \mathrm{MSE}({e}_1, {e}_2, ..., {e}_N, {\hat{e}_1, \hat{e}_2, ..., \hat{e}_N})\ +&\ \mathrm{B}({\hat{e}_1, \hat{e}_2, ..., \hat{e}_N}; \eta).
\end{split}
\end{equation}

We implement the Mean Squared Error (MSE) loss function to quantify the level of discrepancy between the predicted and the actual positions and sizes of GUI elements, represented as

\begin{equation}
\label{eq:equation_mse}
\mathrm{MSE}({e}_1, {e}_2, ..., {e}_N, {\hat{e}_1, \hat{e}_2, ..., \hat{e}_N}) = \dfrac{1}{N}\sum_{i=1}^N|\hat{e}_i - e_i|^2.
\end{equation}

The boundary constraint is used to penalize a predicted element lying outside the interface's screen space:
\begin{equation}
\label{eq:element_boundary_term}
\begin{split}
 \mathrm{B}({\hat{e}_1, \hat{e}_2, ..., \hat{e}_N}; \eta) = & \sum_{t=1}^N\mathrm{B}(\hat{e}_t; \eta) \\
 = &\ \eta \cdot \sum_{t=1}^N\Big(max(-\hat{x}_t, 0) + max(-\hat{y}_t, 0) \\&\qquad \quad + max(\hat{x}_t + \hat{w}_t - w_{\mathrm{UI}}, 0) \\& \qquad \quad + max(\hat{y}_t + \hat{h}_t - h_{\mathrm{UI}}, 0)\Big),
 \end{split}
\end{equation}
where $w_\mathrm{UI}$ and $h_\mathrm{UI}$ are the width and height of the GUI.

We introduce the constraint loss term ($\mathcal{L}_{\mathrm{cons}}$) for estimating the discrepancy between the predicted constraints and the constraints present in the GUI design. This is measured by means of Binary Cross Entropy (BCE), whereby 
evaluates \color{black} a binary decision for each constraint, %\color{red} \sout{denoting} 
namely, \color{black} whether it is satisfied or not, 

\begin{equation}
\label{eq:constraint_loss_term}
\begin{split}
\mathcal{L}_{\mathrm{cons}}(e_1, \ldots, e_N, &\hat{e}_1, \ldots, \hat{e}_N, \mathbf{F},  C) \\= \ &\mathrm{BCE}(C(e_1, \ldots, e_N, \mathbf{F}), ~C(\hat{e}_1, \ldots, \hat{e}_N, \mathbf{F})),
\end{split}
\end{equation}

where $C$ represents the constraints based on the elements and element properties calculated as

\begin{equation}
\mathrm{BCE}(c, \hat{c}) = -c \log(\hat{c}) - (1 - c) \log(1 - \hat{c}).
\end{equation}

The term $-c \log(\hat{c})$ serves to penalize the model when a constraint $c$ that should be satisfied has a predicted probability $\hat{c}$ approaching 0 (where the ground-truth value is 1). This encourages the model to increase the likelihood of satisfying the required constraints. Conversely, the term $-(1 - c) \log(1 - \hat{c})$ punishes the model when the predicted probability $\hat{c}$ is near 1 for a constraint $c$ that should not be satisfied (since the ground-truth value is 0). This term encourages the model to reduce the likelihood of constraints that ought not be satisfied.

As a result, the formulation of the GUI optimization problem is 

\begin{equation}
\label{eq:optimisation_formulation}
\begin{split}
(\hat{e}_1^*, \hat{e}_2^*, ..., \hat{e}_N^*) = \argmin_{\{\hat{e}_1, \hat{e}_2, ..., \hat{e}_N\}} \Big(\dfrac{1}{N}\sum_{i=1}^N(\hat{e}_i - e_i)^2 + \mathrm{B}({\hat{e}_1, \hat{e}_2, ..., \hat{e}_N}; \eta) \\ +~~ \lambda ~~ \mathrm{BCE}(C(e_1, \ldots, e_N, \mathbf{F}), C(\hat{e}_1, \ldots, \hat{e}_N, \mathbf{F}))\Big).
\end{split}
\end{equation}

\section{GUI Representation}

% \begin{figure*}[t]
%   \def\w{0.7\linewidth}
%   \centering
%  \includegraphics[width=\w]{figures/element_node.png}
% \caption{In our graph representation \papername, each GUI element is signified by a separate node with properties. GUI element nodes convey the element properties, including visual appearance, textual content, element type, position, and size.}
%   \label{fig:element_node}
% \end{figure*}

Our proposed method is designed to enrich GUI representation by developing a heterogeneous bipartite graph that covers both GUI element properties and layout constraints, thereby dealing with the intricate arrangement and interrelationships among GUI elements. This graph comprises nodes of two types: GUI element nodes and constraint nodes. The former expresses element properties specific to individual GUI elements, and the latter defines layout constraints for GUI elements in the interface display. Integrating element properties and layout constraints into a single unified graph facilitates a thorough representation of a GUI's elements and layout. While earlier work has integrated element properties into GUI representation~\cite{mapping2020li, pasupat2018mapping, wang2021screen2words, screen2vec2021li}, our approach brings further benefits by not only accounting for the properties of individual GUI elements but also capturing their interrelationships and spatial arrangements within the overall layout.

To ascertain the linkages between GUI elements and constraints, we connect the respective GUI element nodes to the constraint nodes with edges. Specifically, GUI element nodes can only connect to GUI constraint nodes, and \emph{vice versa}. Consequently, the graph constructed represents the GUI structure by establishing the relations between elements and constraints through its edges.

\subsection{Graph Nodes for GUI Elements}
\label{sec:element_node}

In our graph representation \papername, each GUI element is signified by a separate node with properties identifying its position, size, visual appearance, textual content, and type. We encode these properties into an embedding vector and concatenate them to form a single attribute vector for the node (see Figure~\ref{fig:element_and_constraints} a).

\subsubsection{Position Embedding.}
We define a GUI element's position by the coordinates of its top-left and bottom-right points, represented as $(x_1, y_1)$ and $(x_2, y_2)$, respectively. These coordinates specify the element's position and size in the GUI. To represent the position within the graph node, 
a trainable parametric matrix of size $(max(w, h) + 1) \times 256$ is used to encode the position, where $w$ and $h$ are the GUI's width and height, respectively. This matrix maps any coordinate to a 16-dimensional vector. On feeding the four coordinates into this matrix and flattening the resulting embedding, a 64-dimensional vector is output as the position embedding.

\subsubsection{Size Embedding.}
While an element's size can be derived from its position, it is useful to include a size embedding in our representation, especially for tasks such as GUI autocompletion that require a new element to be placed in the GUI at an unknown position. We embed the element size using a process similar to position embedding. This yields a 256-dimensional-vector size embedding.

\subsubsection{Visual Appearance Embedding.}
We encode the visual appearance of GUI elements by extracting high-level features and converting them into a feature vector, which serves as the element's visual representation.

\subsubsection{Textual Content Embedding.}
The textual content of elements is represented by encoding textual information into a vector that captures the semantic meaning and context properties of the text. 

\subsubsection{Element Type Embedding.}
GUI element types are represented as one-hot vectors. For example, in a dataset that contains the three element types Text, List Item, and Button, a button element will be assigned a vector of [0, 0, 1] as its type. In the case of our dataset, which contains 18 distinct element types, the element type embedding is a one-hot vector of length 18, which is processed by a trainable matrix to produce the embedding for the element type.

% \begin{figure*}[t]
%   \def\w{0.9\linewidth}
%   \centering
%  \includegraphics[width=\w]{figures/constraints.png}
% \caption{Constraint nodes express four types of constraint: alignment constraints, same size constraints, element grouping constraints, and multimodal constraints.}
%   \label{fig:constraints}
% \end{figure*}

\subsection{Graph Nodes for Constraints}

Our constructed graph is designed to generalize the definition of constraints. We represent constraints of different types as separate nodes in the graph, which enables easy extension of the graph to include additional kinds of constraints in the future. Currently, we represent four types of constraints as nodes in the graph: alignment constraints, same-size constraints, element grouping constraints, and multimodal constraints (see Figure~\ref{fig:element_and_constraints} b). 

\subsubsection{Alignment Constraint Nodes}
We incorporate alignment constraint nodes into our graph to stipulate the positional affiliation among GUI elements. Each node comprises attributes symbolizing the kind of alignment and the line employed for element alignment. Edges are established between GUI elements and their respective alignment constraint nodes to signify their correlation with alignment. Alignment constraints can express any of six distinct alignments -- namely ``left-aligned,'' ``top-aligned,'' ``right-aligned,'' ``bottom-aligned,'' ``vertical midline-aligned,'' and ``horizontal midline-aligned.'' We employ a one-hot vector to express the alignment type. For instance, a left-alignment type is expressed as $[1, 0, 0, 0, 0, 0]$, indicating left-alignment. We further characterize the alignment line using a two-dimensional vector -- e.g., $[a, 0]$ represents the elements being aligned with $x = a$. We concatenate the alignment type and line representations to produce an eight-dimensional vector.

\subsubsection{Same-Size Constraint Nodes}
To portray GUI elements of the same width or height within the GUI in graphical terms, we devised the notion of uniform size constraint nodes. There are two types of size constraints in our design: identical width and similar height. We consolidate the size type and size value into a single-node attribute vector 
instead of defining them as a one-hot vector. For example, the identical width constraints are defined by $[w, 0]$, where $w$ is the width value, while constraints dictating identical height are defined by $[0, h]$, where $h$ is the height value.

\subsubsection{Element Grouping Constraint Nodes}

Incorporating consideration of related components enhances the structure of GUIs, particularly with regard to elements with comparable functions or belonging to the same category. To depict the element grouping constraints, we define element grouping constraint nodes. We then connect related GUI element nodes to the corresponding grouping constraint node, signifying their inclusion in a particular group.

\subsubsection{Multimodal Grouping Constraint Nodes}
Multimodal grouping constraints enable structured organization of elements of differing types: text, pictures, and other GUI components. 
We create a set of nodes for each multimodal grouping constraint and correlate the relevant elements with their respective nodes. By establishing ties between GUI element nodes and multimodal grouping constraint nodes, we signify placing elements that differ in mode within the same group. In cases of additional multimodal grouping constraints, we create new types of multimodal grouping constraint nodes and repeat the process.

\subsection{Learning GUI Layout Design with Graph Neural Networks}

As Figure~\ref{fig:teaser}a illustrates, we create a graph representation of a GUI layout by organizing GUI element nodes and constraint nodes. Again, these nodes are connected by edges, representing the relationships between elements and constraints. To facilitate GUI design, we can train a graph neural network to take this graph as input and optimize the layout.

\subsubsection{Graph Construction}
The heterogeneous bipartite graph, $\mathcal{G} = (E \cup C, A)$, is constructed from $M$ GUI element nodes and their $N$ corresponding constraint nodes. The former set of nodes is represented by $E = \{e_1, e_2, ..., e_M\}$, and the set of their constraint nodes is denoted by $C = \{c_1, c_2, ..., c_N\}$. In constructing links between element nodes and constraint nodes, we assign the adjacency matrix $A$. 
The value of an element $a_{i,j}$ in that adjacency matrix is set to 1 when the $i$th GUI element is satisfied with the $j$th constraint; otherwise, it is 0.

\subsubsection{Predicting GUI Element Dimensions and Positions}
The primary task of our GNN model is to predict the dimensions and positions of the GUI elements. The predicted GUI element attributes are denoted by $\hat{e}_i = (\hat{x}_i, \hat{y}_i, \hat{w}_i, \hat{h}_i)$.
These predictions are produced through parameterized functions specific to the GNN model. The model parameters are denoted by $\theta$, and the model that generates the GUI element predictions is denoted as $\textrm{GNN}_{\theta}$. The predicted GUI elements $\hat{e}_i$ are then computed as

\begin{equation}
\hat{e}_i = \textrm{GNN}_{\theta}(\mathcal{G}).
\end{equation}

\subsubsection{Optimization of GNN Parameters}
The GNN model's $\theta$ parameters are optimized by minimizing the previously defined objective function $\mathcal{L}$; see Subsection \ref{sec:objective_function}. The optimization process can be represented as

\begin{equation}
\theta^{*} = \argmin_{\theta} \mathcal{L}(\hat{e}_1, \hat{e}_2, ..., \hat{e}_N, \mathbf{F}; \theta),
\end{equation}

where $\theta^{*}$ denotes the optimized parameters of the GNN model.
After optimizing the GNN model's parameters, $\theta^{*}$, we can use this model to design new GUI layouts. For a new design, the trained GNN takes the graph-form representation of GUI elements and constraints as input, and then outputs its predicted dimensions and positions for GUI elements.
Our approach makes it easy to incorporate new design constraints by modifying the construction of the graph or the objective function.

\section{GUI Autocompletion}

\label{sec:autocompletion}

\begin{figure*}[t!]
% \marginparsep=30pt
% \marginnote{\color{violet}
% R2: We add Fig 3 and revise Sec 5.1 to better explain the GNN model of autocompletion. \\~\\}
  \def\w{\linewidth}
  \centering
 \includegraphics[width=\w]{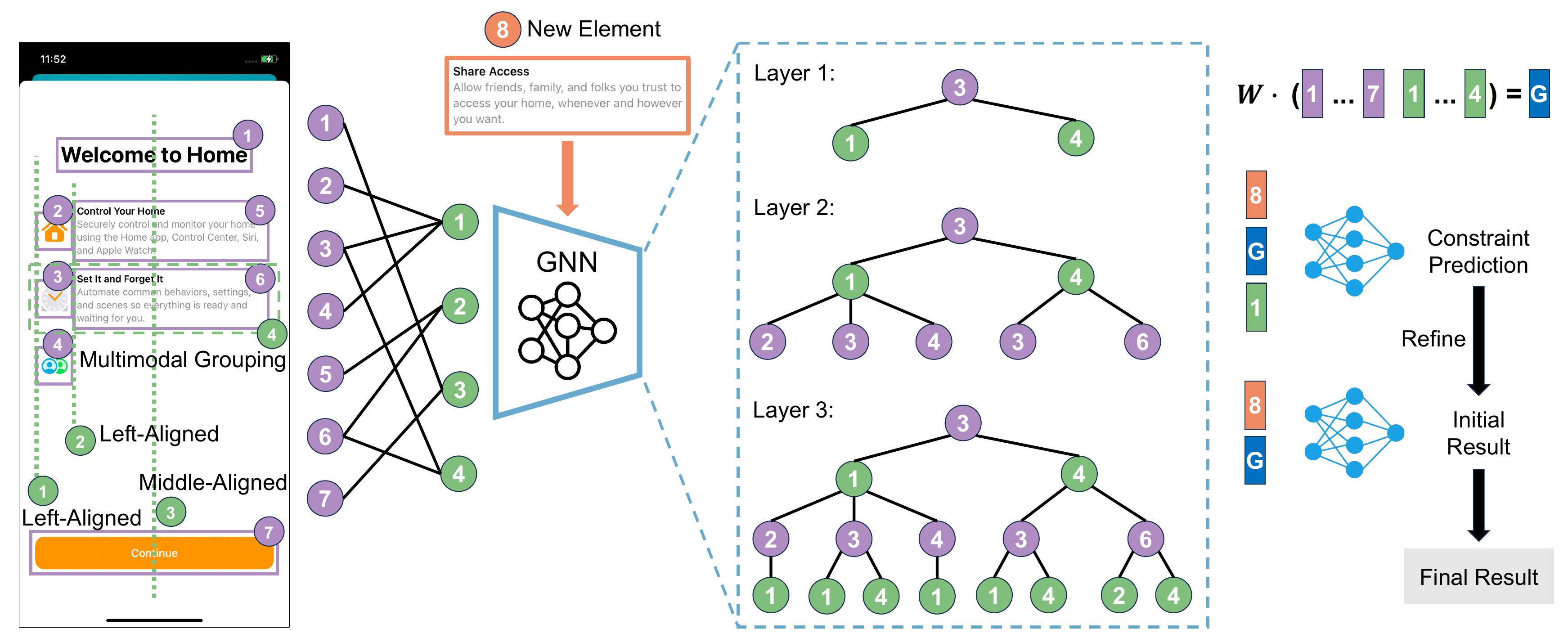}
\caption{\add{\papername~ was adapted for the autocompletion task: We first encode the graph representation of the GUI via the GNN. We only illustrate some parts of the graph for simplicity. Element 8 is the target to-be-placed element.
In each GNN layer, nodes perform aggregation from their respective neighbors. To illustrate, consider element node 3. As it goes through the GNN layers, it accumulates information from related constraint nodes and other element nodes. This process results in feature embedding vectors for all nodes, including both element nodes and constraint nodes within the graph.
We compute the graph embedding as a weighted average of the node embeddings with the weight matrix $W$. We then concatenate the target element's embedding vector, the graph embedding, and a constraint embedding and send it to fully connected layers to predict whether the target to-be-placed element should satisfy the constraint. Simultaneously, we concatenate the target element's embedding and the graph embedding to predict the initial position and size of the target element. Integrating these predictions with the constraints, we subsequently refine the position and size to obtain the final results.}}
\Description{This figure shows the process of autocompletion. Given the graph for the GUI, it goes through a graph neural network. Then we predict the final results by refining the initial results by the predicted constraints.}
  \label{fig:graph_prediction}
\end{figure*}

\marginparsep=30pt

To demonstrate the utility of our graph representation, we propose an autocompletion method that uses our representation approach to enable interactive iterative design. 
GUI autocompletion is challenging due to the computational complexity involved in accurately predicting suitable GUI elements.
Given fixed screen dimensions, our method automatically generates suggestions for finishing a partially completed GUI layout by iteratively predicting the positions of remaining unplaced GUI elements. 
\add{Our method suggests position, size, and confidence level for each unplaced element based on the partial GUI. It enables designers to receive suggestions when they complete the design of each element, without the need to predefine all GUI elements beforehand. Moreover, if designers have additional unplaced GUI elements ready, our method iterates over each, providing suggestions for their positions, sizes, and confidence levels.}
It can significantly reduce the manual effort required for design.
Autocompletion that produces high-quality GUIs is rendered difficult by the high computational complexity of evaluating all possible combinations of GUI elements.

Prior studies have explored the autocompletion task; however, they were only capable of handling wireframes. Li et al.~\cite{li2020auto} used the GUI layout hierarchy to perform GUI autocompletion. Although the hierarchy does capture the structure of the layout, it accounts for only the grouping and containment relationships between GUI elements. It neglects the alignments and relative sizes, which are important layout constraints. On the other hand, Br{\"u}ckner et al.~\cite{bruckner2022learning} proposed a method of constructing a graph with reference to differences in position between GUI elements. However, it does not consider the properties of GUI elements.
GRIDS~\cite{dayama2020grids} is a grid-layout-based optimization approach for autocompletion considering constraints such as alignment and grouping. With integer programming, GRIDS produces results by searching for optimal available placements for unplaced elements. 
Our method, considering both element properties and constraints, fills the gap, crossing the void to generate more desirable predictions.

\subsection{Target GUI Element Prediction}

Given a partial GUI, we set out to predict both the size and the position of a target element (with a fixed aspect ratio) and the associated constraints it should follow. \add{As shown in \autoref{fig:graph_prediction}}, the process begins with constructing a graph representation of the partial GUI, denoted as $\mathcal{G}_p$. 
Exploiting GNNs, we encode this graph to yield feature vectors for all the nodes, including element nodes and constraint nodes within the graph. 
\add{Within each GNN layer, nodes are aggregated from their respective neighbors. Going through the GNN layers, each node iteratively accumulates information from its associated constraint nodes and other element nodes.}

\begin{equation}
\begin{split}
\mathcal{G}_p \xrightarrow{\mathbf{GNN}} \{\mathbf{h}_{ele,1}, \mathbf{h}_{ele,2}, ...\}, \{\mathbf{h}_{align,1}, \mathbf{h}_{align,2}, ...\}, \\ 
\{\mathbf{h}_{size,1},  \mathbf{h}_{size,2}, ...\}, \{\mathbf{h}_{eg,1}, \mathbf{h}_{eg,2}, ...\}, \{\mathbf{h}_{mg,1}, \mathbf{h}_{mg,2}, ...\},
\end{split}
\label{}
\end{equation}

where  $\{h_{ele,1}$, $h_{ele,2}$, ...$\}$ are the feature vectors for element nodes, $\{h_{align,1}$, $h_{align,2}$, ...$\}$ those for alignment constraint nodes, $\{h_{size,1}, h_{size,2}, ...\}$ those for size constraint nodes, $\{h_{eg,1}, h_{eg,2}, ...\}$ the ones for element grouping constraint nodes, and $\{h_{mg,1}, h_{mg,2}, ...\}$ the vectors for multimodal grouping constraint nodes. 

After this, the feature vector $h_{\mathcal{G}_p}$ for the entire graph $\mathcal{G}_p$ is computed. This vector is obtained by way of the weighted summation of the average feature vectors of each node type. The weight matrices $\mathbf{W}{ele}$, $\mathbf{W}{align}$, $\mathbf{W}{size}$, $\mathbf{W}{eg}$, and $\mathbf{W}_{mg}$ are trained alongside GNN parameters, making sure an end-to-end training process ensues that eliminates manual selection: % FYI: Avoiding `ensure', which is UKish. -als

\begin{equation}
\begin{split}
h_\mathcal{G} = &\mathbf{W}_{ele} \cdot \mathrm{avg}\{\mathbf{h}_{ele,1}, \mathbf{h}_{ele,2}, ...\} + \mathbf{W}_{align} \cdot \mathrm{avg}\{\mathbf{h}_{align,1}, \mathbf{h}_{align,2}, ...\} \\ &+ \mathbf{W}_{size} \cdot \mathrm{avg}\{\mathbf{h}_{size,1}, \mathbf{h}_{size,2}, ...\} + \mathbf{W}_{eg} \cdot \mathrm{avg}\{\mathbf{h}_{eg,1}, \mathbf{h}_{eg,2}, ...\} \\ &+ \mathbf{W}_{mg} \cdot \mathrm{avg}\{\mathbf{h}_{mg,1}, \mathbf{h}_{mg,2}, ...\}.
\end{split}
\label{eq:graph_embedding}
\end{equation}

The resulting graph feature vector and the embedding of the target element are concatenated and fed into fully connected layers. This facilitates position and size predictions for the target GUI element $\hat{e}_t = (\hat{x}_t, \hat{y}_t, \hat{w}_t, \hat{h}_t)$.
Furthermore, our approach extends to predicting the constraints that should be satisfied by the target GUI element. We utilize the objective function outlined in Section \ref{objective} to perform fine optimization of the target element's positioning, dimensions, and adherence to constraints. \add{For each constraint within the partial GUI, we concatenate the graph feature vector, the embedding of the target element, and the specific constraint's feature vector. This concatenated vector is propagated through fully connected layers to predict the probability of each constraint required for the target GUI element's satisfaction. Simultaneously, we concatenate the target element's embedding and the graph embedding to predict the initial position and size of the target element. Integrating these predictions with the constraints, we subsequently refine the position and size to obtain the final results.} 

\subsection{Confidence Levels}

To guide the process of ascertaining the level of confidence in the outcomes, below we describe how we compute confidence levels such that we can avoid offering potentially questionable predictions.
Conveying the confidence level – whether it is low, medium, or high – of a certain prediction enables software tools and designers to take suitably informed actions. An application of this feature will be discussed later in the designer study.

%we propose to that checks the predicted size and position of the model-generated target element, denoted as $\hat{e}_t = (\hat{x}_t, \hat{y}_t, \hat{w}_t, \hat{h}_t)$, against the predicted constraints, $\hat{c}$. 
%Below, we describe how to compute confidence levels such that we can avoid offering potentially questionable predictions.

\subsubsection{High Confidence}

High confidence is validated in terms of alignment and uniform size constraints.
When the disparity between the predicted alignment line and the position of the target element falls below the threshold value $\sigma$, % Yes? -als
refinement is performed by aligning the position with the projected alignment line. For instance, if the predicted constraint entails left-alignment with the line $x = a$, and if $|\hat{x}_t - a| \leq \sigma$, the $\hat{x}_t$ value is adjusted to match $a$. This threshold was set to $\sigma = 20$ pixels in our experiments.
Likewise, when the difference between the size of the target element and the sizes of other elements, as predicted by constraints promoting uniformity, is below % Yes? -als
the $\sigma$ limit, the size is adjusted to match the uniform size value. For example, if the projected constraint dictates that the target element should possess the same width as elements with a width of $b$, and if $|\hat{w}_t - b| \leq \sigma$, the $\hat{w}_t$ value is adjusted to align with $b$.
When both $(\hat{x}_t, \hat{y}_t)$ and at least one of $\hat{w}_t$ and $\hat{h}_t$ can be verified, we assign a \textit{high} confidence level to the outcome, since the fixed aspect ratio of the target element allows deducing the remaining attribute.

\subsubsection{Medium Confidence}

Medium confidence is established via element and multimodal grouping constraints.
In scenarios wherein confirmation is unattainable for $(\hat{x}_t, \hat{y}_t)$ and at least one of $\hat{w}_t$ and $\hat{h}_t$, grouping constraints come into play. These constraints, encompassing both element and multimodal grouping, reveal patterns among elements and can be exploited for refinement of positions and sizes. For example, vertical groupings often entail consistent widths and equidistant spacing between elements vertically. Consequently, if $|\hat{w}_t - \mathrm{avg}(w_i)| \leq \sigma$, where $w_i$ represents widths of other elements in the target's vertical group, $\hat{w}_t$ is adjusted to match $\mathrm{avg}(w_i)$. If $(\hat{w}t - w_l) - \mathrm{avg}(|w_i - w{i-1}|) \leq \sigma$, the $\hat{w}t$ value is set to $w_l + \mathrm{avg}(|w_i - w{i-1}|)$. In instances where $(\hat{x}_t, \hat{y}_t)$, along with at least one of $\hat{w}_t$ and $\hat{h}_t$, can be verified, the result gets accorded a \textit{medium} confidence rating.

\subsubsection{Low Confidence}
In all other cases, the outcomes are assigned a \textit{low} confidence rating.

\section{Experiments For Autocompletion}

We conducted experiments for the autocompletion task to show the effectiveness of our representation.
We created a dataset with partial GUIs and then evaluated our method's prediction quality through qualitative and quantitative experiments. Additionally, we conducted an ablation study to demonstrate the necessity of each constraint type taken into account.

\subsection{Dataset and Training Process}
\label{sec:dataset}
For the evaluation, we took the ENRICO dataset~\cite{leiva2020enrico}, a subset of the RICO dataset~\cite{deka2017rico} including cleaner mobile GUI information and the VINS~\cite{bunian2021vins} GUI dataset, as our basis for creating a mobile dataset for GUI autocompletion.
We improved the dataset's quality through several steps. Initially, we excluded layouts in the dataset that contain three or fewer.
GUI elements. After this, we employed the UIED model~\cite{xie2020uied} to enhance the precision of element types and refine the bounding boxes of the GUI elements. We then made further adjustments manually to correct the bounding boxes of the elements. Our refined dataset contains 5,653 GUIs in total.

To evaluate our model's performance, we followed a fivefold cross-validation approach; this technique involved partitioning the GUI dataset into five equal-sized folds. Four of the folds, with approximately 4,522 GUIs, served for training our model, while we reserved one fold, encompassing around 1,131 GUIs, for testing.
In our experiments, each fold was utilized once for testing, with the remaining four folds serving as the training data.
Using fivefold cross-validation helps to validate the generalization of the model to unseen data and offers a more comprehensive view of the model's behavior by averaging its performance across multiple test sets, thus reducing the impact of random variations in the data. 

To create our dataset for GUI autocompletion, for each GUI, we randomly kept a chunk of GUI elements on the display. We removed other GUI elements to create a partial GUI and store the potential ``next GUI element'' to be added for completing the given partial GUI. By this mechanism, we obtained a partially completed GUI with missing elements and a target element that we need to predict, given the partially completed GUI. Note that each partial GUI often had more than one potential target GUI element. With this method, we can generate various partial GUIs and corresponding target elements. We generated partial GUIs for training from the complete GUIs in the training data, doing similarly for the test data. In total, each fold of complete GUIs yielded approximately 171,212 pairs of partial GUIs and corresponding target elements for training. Consequently, for each experiment, we used a training dataset containing about 684,849 incomplete-GUI--target pairs and a test dataset comprising approximately 171,212 pairs. 

\subsection{Implementation Details}

\subsubsection{Embeddings}
We encoded the visual appearance of the element by using a pre-trained ResNet152 model~\cite{he2016deep}. Through this model, which is able to extract high-level features from images, we generated a feature vector that represents the element's visual appearance.
For encoding the textual content of the GUI elements, we used a pre-trained BERT model~\cite{devlin2018bert}. A Transformer-based 
neural-network architecture pre-trained on a large corpus of text data, BERT can generate a 768-dimensional vector representing the text, which we applied to extract features from the interface elements.
In a technique that improves the efficiency of models utilizing this representation, we introduced an ``unknown'' token for infrequent words. Infrequent words are often inadequately represented in training data. That can lead to overfitting. Incorporating an unknown token enables the model to generalize its predictions for previously unseen words and simplify the representation to facilitate the model's processing. 
To implement this approach, we began by computing the frequency of each text element. 
If the text occurred fewer than three times, we replaced it with the special token [UNK] in BERT, representing an unknown word. We then used BERT to generate the text content embedding.

\subsubsection{Graph Neural Networks}

We applied the SAGEConv model \cite{hamilton2017inductive}, a variety of GNN models that is suited to training heterogeneous graphs. 
The SAGEConv model employs a message-passing technique to propagate information through the graph to convey it from a node's neighborhood to the node itself, thereby improving its feature representation. SAGEConv can capture the relationships between nodes of different types in the graph. The output feature vectors are 256-dimensional $h\in\mathbb{R}^{256}$. The trainable weights for computing $\mathbf{W}_{ele}$, $\mathbf{W}_{align}$, $\mathbf{W}_{size}$, $\mathbf{W}_{eg}$, $\mathbf{W}_{mg}\in\mathbb{R}$ are in the dimension of ${256\times 256}$. 

\begin{figure*}[!]
  \def\w{0.98\linewidth}
  \centering
 \includegraphics[width=\w, trim=0 0 0 0]{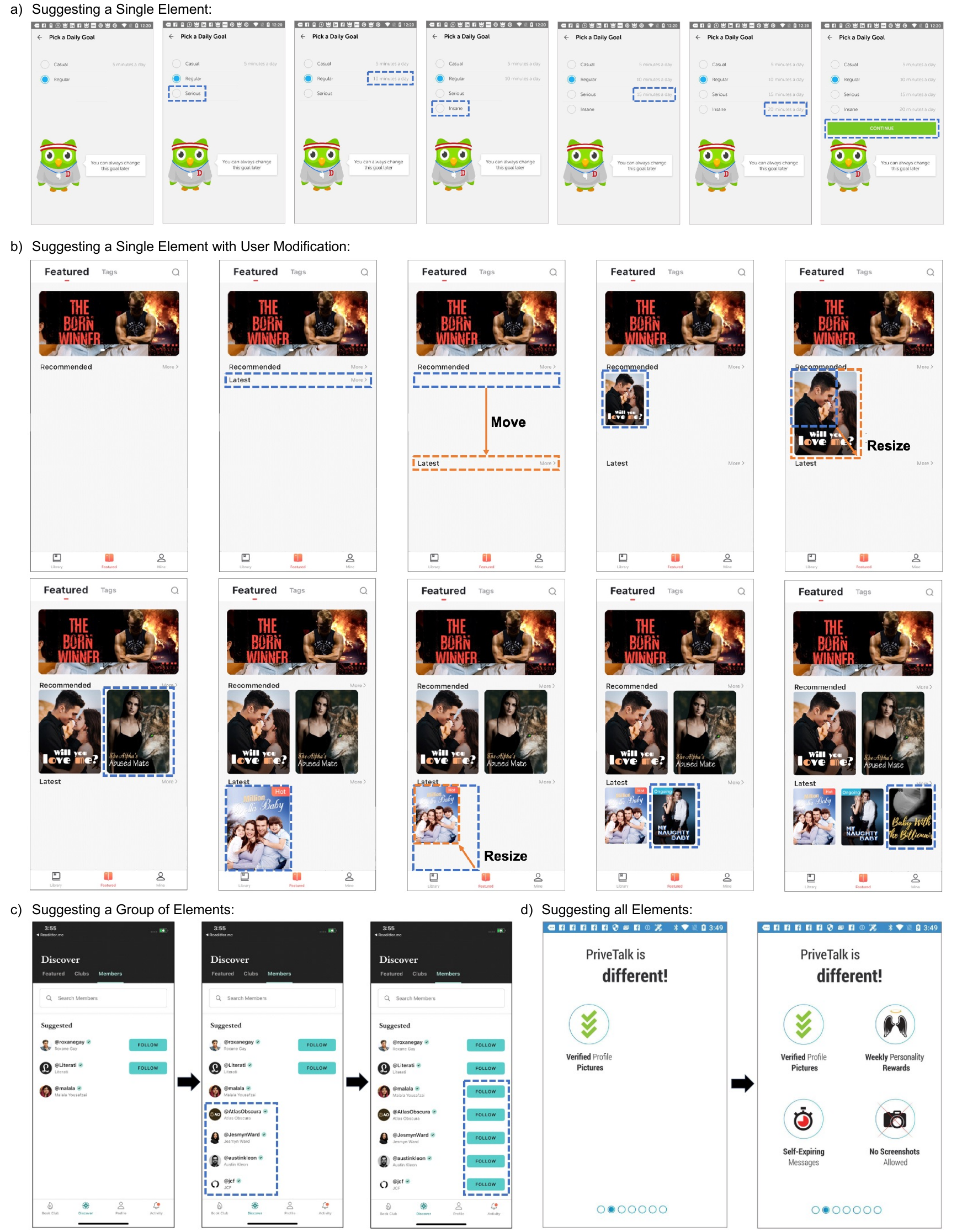}
\caption{a) Our model can iteratively predict unplaced GUI elements (shown in blue bounding boxes).
b) Designers can make adjustments (orange), including moving, resizing, or re-selecting GUI elements.
c) The model's capability to predict groupings allows for the placement of elements together as a group.
d) The model can also predict all the elements simultaneously.
}
\Description{This figure shows that our model can a) suggest a single element; b) suggest a single element with user modification; c) suggest a group of elements; and d) suggest all elements.}
  \label{fig:new_results}
\end{figure*}

% \begin{figure*}[!]
%   \def\w{0.93\linewidth}
%   \centering

%  \includegraphics[width=\w, trim=0 40 0 25]{figures/results2.pdf}
% \caption{ 
% Results in the GUI autocompletion task from iteratively predicting unplaced GUI elements (in blue bounding boxes). We allowed designers to make adjustments in accordance with their preferences (orange), which might include moving, resizing, or re-selecting GUI elements. The underlying graph representation gets updated accordingly and informs subsequent predictions. 
% }
%   \label{fig:results2}
% \end{figure*}

\begin{ADD}

% \begin{figure*}[!]
%   \def\w{\linewidth}
%   \centering
%  \includegraphics[width=\w, trim=0 40 0 25]{figures/suggestions.png}
% \caption{ \add{In addition to the iterative prediction of elements, we provide more options in element suggestion. a) Facilitating the suggestion of a group of elements by utilizing the model's capability to predict groupings, thereby expediting the design process. b) Predicting all elements simultaneously provides a complete overview of the final layout, yet it poses challenges when making adjustments after the initial predictions.}
% }
%   \label{fig:results3}
% \end{figure*}

%\subsection{Element Suggestion Options}

% \marginparsep=30pt
% \marginnote{\color{violet}
% 1AC, 2AC, R2, R3: We provide multiple element suggestion options as more efficient interaction design methods. (Added Fig 6 and Sec 6.3)  \\~\\}

\subsection{Qualitative Evaluation}

\add{To enhance the usability of the model for designers, we introduce three types of element suggestion options and show the results.}

\subsubsection{Suggesting a Single Element} 

As Figure~\ref{fig:teaser}b indicates, iteratively predicting the sizes and positions of yet-unplaced elements by means of the updated graph representation helps support designers by autocompleting partially completed designs. By default, we loop over all elements still to be placed and select the one with the highest associated confidence level to add (\autoref{fig:new_results}a). 
Furthermore, our setting allowed designer-in-the-loop interaction wherein designers can make adjustments to the GUI design as their preferences dictate after every iteration. 
They could move the element, resize it, or 
select an alternative GUI element for placement. The changes are visible immediately, and the underlying graph representation gets updated accordingly, so that subsequent predictions can work from it, as demonstrated in Figure~\ref{fig:new_results}b.

% the model suggests each element in the order of confidence level for multiple unplaced GUI elements. As shown in~\autoref{fig:new_results}b, predicting one element at a time gives designers more control over GUI design, allowing adjustments like moving, resizing, or re-selecting GUI elements. The underlying graph representation gets updated accordingly and informs subsequent predictions.

\subsubsection{Suggesting a Group of Elements}

Our model predicts grouping constraints for each element based on the partial GUI. As shown in \autoref{fig:new_results}c, if multiple elements share the same grouping constraint, our model suggests these grouped elements together, thereby expediting the prediction process.

\subsubsection{Suggesting All Elements}

Alternatively, in~\autoref{fig:new_results}d, the model can predict all elements simultaneously. The final results iterate over each element, placing those with the highest confidence levels first, based on the updated partial GUI. While providing a complete view, modifying results can be more challenging compared to adjustments in the iterative prediction process.

\end{ADD}

\begin{figure*}[t]
\def\w{0.24\linewidth}
 \centering
\begin{tabular}{*3c}
\includegraphics[height=\w, trim=35 30 35 10]{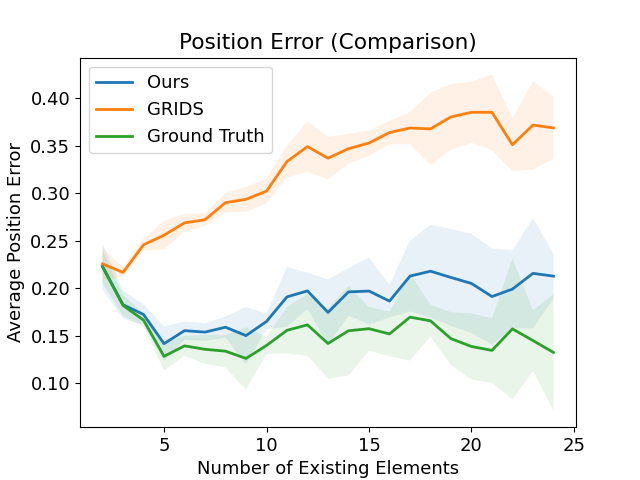} &
\includegraphics[height=\w, trim=35 30 35 10]{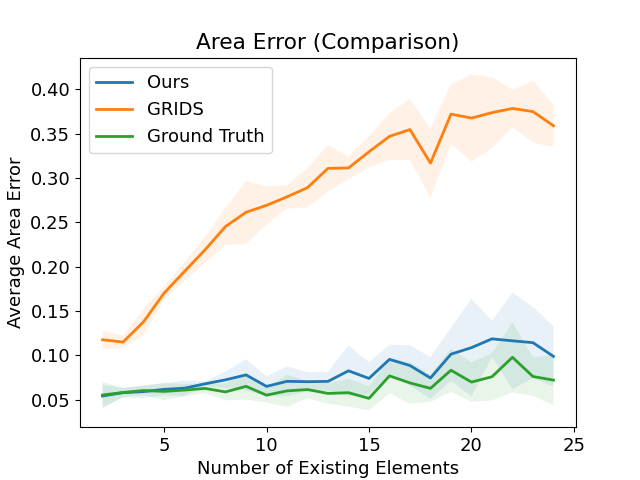} &
\includegraphics[height=\w, trim=35 30 35 10]{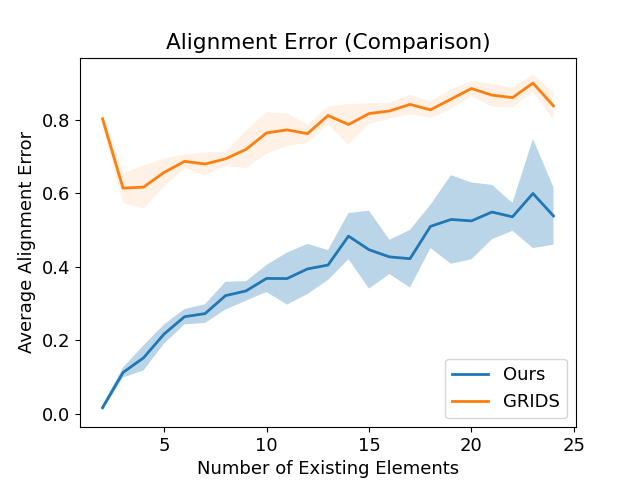}
\end{tabular}
\caption{Comparison of our model with GRIDS~\cite{dayama2020grids}, an autocompletion approach using integer programming, and the established upper bound for the model's performance exploiting ground-truth constraints to predict positions and sizes. The evaluation used three metrics: position error, size error, and alignment error. This comparison incorporates fivefold cross-validation to assure reliability, with the mean and standard deviation illustrated in the corresponding plots.
}
\Description{Plots about the comparison to GRIDS on position error, area error, and alignment error.}
\label{fig:comparison}
\end{figure*}

\begin{figure*}[t]
\def\w{0.24\linewidth}
 \centering
\begin{tabular}{*3c}
\includegraphics[height=\w, trim=35 30 35 10]{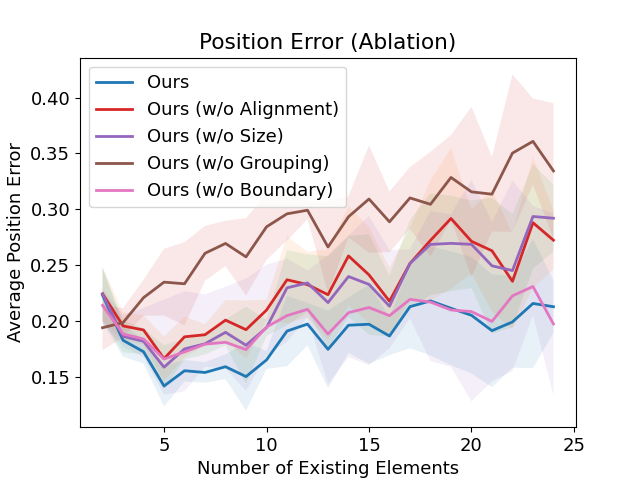} &
\includegraphics[height=\w, trim=35 30 35 10]{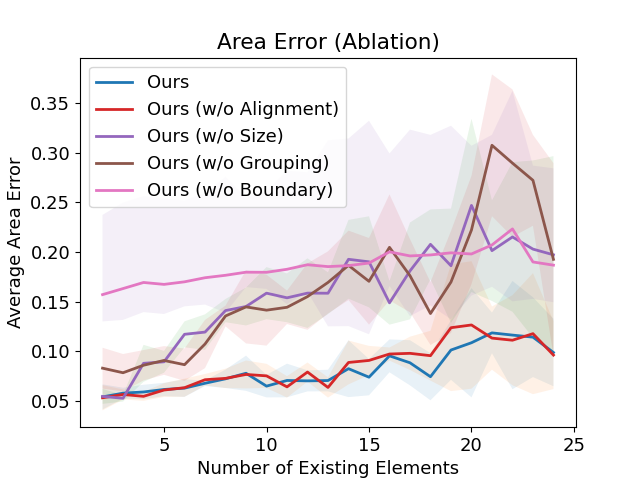} &
\includegraphics[height=\w, trim=35 30 35 10]{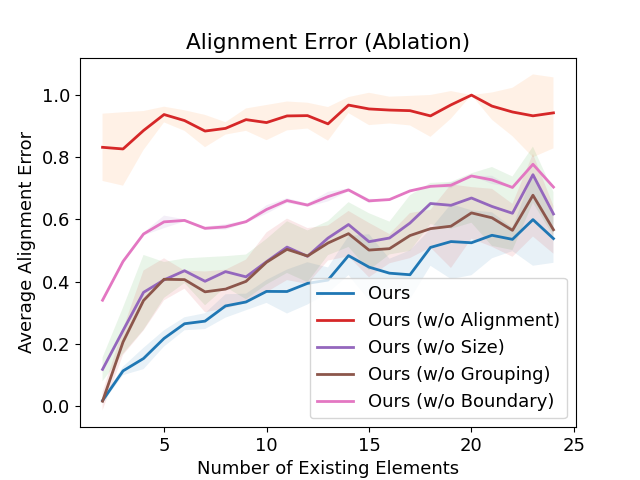}
\end{tabular}
\caption{Results from an ablation study comparing our model's performance to ablated models in which each type of constraint has been removed.
}
\Description{Plots about the ablation study on position error, area error, and alignment error.}
\label{fig:ablation}
\end{figure*}

\subsection{Quantitative Evaluation}
\label{sec:metrics}

To evaluate the accuracy of our autocompletion approach, we assessed its single-step prediction by three metrics. For this purpose, we denoted the top-left point of the predicted GUI element as $(\hat{x}, \hat{y})$ and its corresponding ground truth as $(x, y)$. Similarly, we denoted the predicted size of the target GUI element as $(\hat{w}, \hat{h})$ and its corresponding ground truth as $(w, h)$. Finally, $w_\mathrm{UI}$ and $h_\mathrm{UI}$ represent the width and height of the user interface.

\subsubsection{Metrics} We established separate metrics for assessing the predictions' accuracy with regard to position, size, and alignment. All three metrics have a range between 0 and 1, where a lower value indicates a better prediction.

\begin{itemize}
  \item \textbf{Position Error (PosError):} 
    The PosError metric measures the relative distance between the predicted position of the GUI element and the corresponding ground-truth position. Calculating the error entails ascertaining the distance between the predicted and ground-truth positions, then normalizing it by the maximum possible distance that the element can move,
\begin{flalign}
\loss_{\textrm{PosError}}=\dfrac{||(\hat{x}, \hat{y}) - (x, y)||_2}{\sqrt{(w_\mathrm{UI} - \hat{w})^2 + (h_\mathrm{UI} - \hat{h})^2}}.
\end{flalign}

  \item \textbf{Area Error (AreaError):}  
    The AreaError metric evaluates the difference between the predicted size of the GUI element and the corresponding ground-truth size. The difference between the predicted and ground-truth sizes is normalized in terms of the maximum size between the predicted size and the ground-truth size,
\begin{flalign}
\loss_{\textrm{AreaError}}=\dfrac{|\hat{w}\cdot \hat{h} - w \cdot h|}{\max(\hat{w}\cdot \hat{h}, w \cdot h)}.
\end{flalign}

  \item \textbf{Alignment Error (AlignError):} 
   The AlignError metric judges the proportion of the alignments predicted correctly. This figure is calculated by dividing the number of correctly predicted alignments for the target element by the total number of alignments that the predicted element should satisfy. 
\end{itemize} 

\subsubsection{Comparison}

% \marginparsep=30pt
% \marginnote{\color{violet}
% 2AC, R3: We clarify the reason we chose GRIDS to compare)  \\~\\}
\add{We compared our model and GRIDS~\cite{dayama2020grids}, an optimization approach based on grid layout designed for autocompletion while considering constraints, including alignment, element location, rectangular outline, and preferred element positions. We chose GRIDS for comparison due to the following reasons: 1) it also takes constraints into account, 2) the majority of existing GUI presentations do not perform autocompletion, and enabling autocompletion is non-trivial, 3) other relevant prior works~\cite{li2020auto, bruckner2022learning} addressing autocompletion tasks are not open-sourced.}
With integer programming, GRIDS produces results by seeking the optimal placements available for unplaced elements. It generates multiple optimized solutions, each accompanied by a confidence value. We chose the solution with the highest confidence value among each model's first 10 solutions.  
In addition, we established an upper bound for the model's performance by employing ground-truth constraints to predict element positions and sizes. Because of the ambiguity of GUI element placement, the ground truth thus defined was not assigned a zero-loss value. The ambiguity arises from the fact that some elements do not satisfy enough constraints; that issue, in turn, makes accurate prediction of their placement challenging. Our comparison by all three metrics was performed relative to the number of preexisting elements in the partial GUIs, as Figure~\ref{fig:comparison} illustrates. We used fivefold cross-validation for the comparisons (our plots present both mean and standard deviation values), with a sample of 10,000 test data from each fold used for model evaluation. Since our technique uses ground-truth alignments in predicting the ground truth, we do not have ground-truth results for alignment error. 
The results show that our model predicts more accurate positions and sizes, with more accurate alignments, than GRIDS. 
Finally, we computed the inference time needed by the models. Our model performs one-step predictions in 0.148 seconds, on average, with our test data on a single RTX4090 GPU, while GRIDS takes much longer, at 67.4 seconds.

\subsubsection{The Ablation Study}

Our ablation study compared the proposed model to ablated models, each lacking one specific type of constraint. This ablation study, the results of which are depicted in Figure~\ref{fig:ablation}, showed the necessity of each constraint for our model's performance.

\section{Comparison Study}

% \marginparsep=30pt
% \marginnote{\color{violet}
% 2AC: We clarify that the test images are from the same test dataset as in the previous section)  \\~\\}
We performed a comparison study to evaluate our model against GRIDS~\cite{dayama2020grids} for one-step prediction. 
 \add{We got test images randomly sampled from the test dataset described in \autoref{sec:dataset}. All the images are mobile GUIs.}
As Graph4GUI optimizes positions and sizes with content and graphics, while GRIDS only handles wireframe layouts, the comparison focuses on wireframes.
To ensure fairness, we trained our model on wireframe GUIs without visual appearance, textual content, or element type, showcasing higher perceived quality despite not fully utilizing Graph4GUI's capabilities.

\begin{figure*}[t!]
  \def\w{0.7\linewidth}
  \centering
 \includegraphics[width=\w]{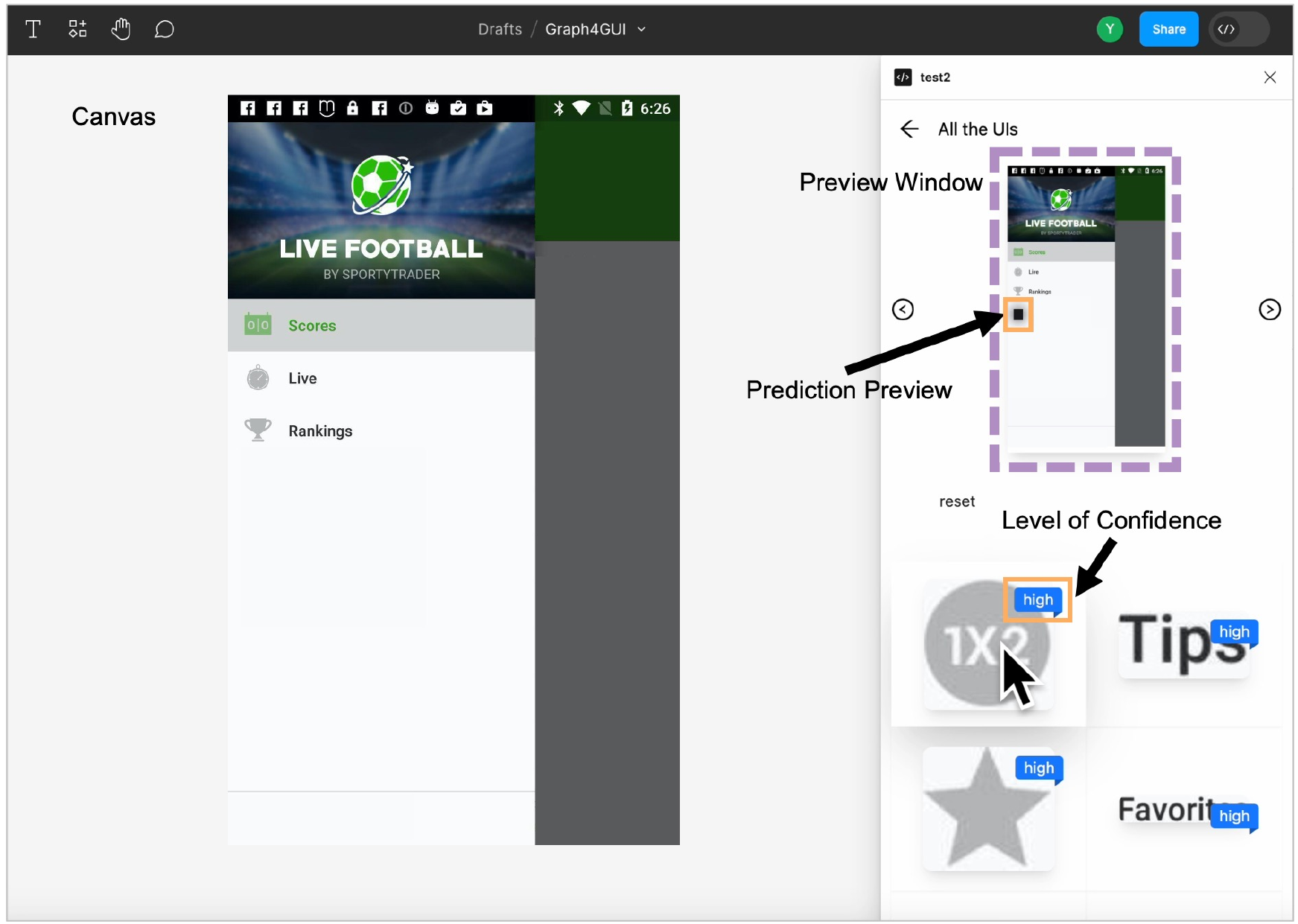}
\caption{We implemented our method as a Figma plug-in.
The plug-in offers GUI element prediction suggestions with confidence levels, helping designers prioritize element selection. It also features a preview window that shows a prediction preview of the element placed on the GUI when hovering over an element, providing an intuitive view to help decision-making. After selecting a GUI element, the plug-in can automatically place it in the suggested position and size on the canvas.}
\Description{This figure shows the appearance of our plug-in.}
  \label{fig:plugin}
\end{figure*}

\subsection{Method}

\subsubsection{Participants}

We enlisted 35 participants (25 female, 9 male, 1 other) through social media, averaging 26.49 years (SD = 3.07). All had normal or corrected-to-normal vision, and none were colorblind. Local regulations did not mandate formal ethics review.

\subsubsection{Experimental Design}

From 1,000 randomly sampled partial GUIs with a to-be-placed element, we randomly selected 100 presented to each participant for comparison between our method and the GRIDS method.

\subsubsection{Apparatus}

 Pairs of GUI images were presented side by side on a custom webpage in randomized order.

\subsubsection{Procedure}

After completing a demographics questionnaire, participants viewed GUI pairs and selected the preferred one based on personal criteria like design, layout, or aesthetics. Preferences were indicated by choosing the left or right image, or ``They are equally good''. Participants could assess up to 100 pairs, stopping at their discretion.

\subsection{Findings}

We received responses for 3,367 image pairs from 35 participants. Preferences were as follows: 456 for GRIDS (13.54\%), 2,368 for our model (70.33\%), and 543 expressing no preference (16.13\%). 
The difference between our method and GRIDS was statistically significant ($\chi^2 = 3115.8, p < .001$).
This finding attests that our model indeed produces more visually appealing suggestions that show better alignment than the baseline method's output, backing up our conclusions with evidence from participants' preferences.

 \section{Designer Study}

We conducted a user study to evaluate the effectiveness and usability of our method for assisting with GUI autocompletion. The aim was to assess both the impact on design efficiency and the subjective experience. 
To guide the design of the study, we established the following objectives:

\begin{enumerate}
\item Determine whether our technique enables designers to enhance the efficiency of the design process.
\item Evaluate the quality of suggestions provided by our model.
\item Explore how designers utilize each function of the tool, specifically the element prediction preview, the element prediction, and the confidence rating for the predictions.
\item Ascertain whether designers perceived our technique as helpful for their GUI design practice.
\end{enumerate}

\subsection{Method}

\subsubsection{Plug-in}

Our method is implemented as a Figma plug-in (Figure~\ref{fig:plugin}), offering GUI element predictions with confidence levels. Given a partially completed GUI and a list of elements to be placed, the plug-in computes confidence levels for each prediction, aiding designers in prioritizing placements. The plug-in includes a preview window that displays a wireframe version of the GUI when hovering over an element, allowing designers to assess results intuitively. Upon selecting a GUI element, the plug-in automatically positions and sizes it on the canvas.

\subsubsection{Participants}

Six GUI designers with diverse experience levels were recruited through email lists, local networks, and social media platforms. Participants, aged 21 to 33 (mean age 26.5), were either UI/UX designers or HCI/design students, all experienced in Figma for GUI design. The group comprised four females and two males. Prior to the study, participants received full information about the conditions and gave informed consent. Local regulations do not require formal ethics review.

%We recruited six GUI designers, with varying levels of experience. Participants were obtained through email lists, local networks, and social-media platforms. The age range of the participants was 21 to 33, with a mean age of 26.5. All participants were either user-interface/user-experience designers with companies or HCI/design students, and they all had experience in using Figma for GUI design. The gender breakdown was four female participants and two male participants.
%Before participating in the study, all participants were fully informed about the study's conditions and provided their informed consent. Local regulations do not require formal ethics review.
% 23 28 29 32 21 26

\subsubsection{Materials}

Participants used the lab's laptops for various activities, including interacting with the Figma interface using our plug-in and filling out a questionnaire. The study involved a practice task to acquaint participants with Figma and our plug-in, and six GUI design tasks (three with our plug-in and three without). Each task provided a brief outlining the GUI design purpose, a GUI to be completed, and a list of elements to place.

%The participants utilized the lab's laptop computers to engage in several activities, among which were using the Figma interface with our plug-in and completing a questionnaire. The study consisted of two design tasks: a practice task, aimed at familiarizing participants with Figma and our plug-in, and three GUI design tasks for each of the two modes (with our plug-in and without), featuring realistic design cases. Each design task included a brief outlining the GUI design purpose and a GUI layout that needed to be completed with a list of to-be-placed elements. 

\subsubsection{Experiment Design}
The study used a within-subject design, exposing participants to two conditions. In one condition, they freely used our plug-in with Figma's standard features to complete three GUIs (login, shopping, and menu pages). In the baseline condition, participants completed the same GUIs without the plug-in. Task and condition order were fully counterbalanced to eliminate any potential bias.

\subsubsection{Procedure}

After signing the consent form, participants completed a design background questionnaire and underwent a Figma tutorial. They were then tasked with creating six GUIs in two conditions: with and without our plug-in. After each task, participants rated their designs and assessed the perceived task load. In the with-plug-in condition, they also evaluated the plug-in's specific features. We used the System Usability Scale (SUS) for overall usability assessment. We conducted final interviews for participants to compare their experiences, provide feedback on plug-in features, and identify weaknesses in our tool.

\subsection{Quantitative Findings}

Our tool's usability and helpfulness were quantitatively evaluated using the System Usability Score, participant ratings of features and resulting GUIs, and task completion times.

\subsubsection{System Usability Score} 

Following established practices~\cite{brooke1996sus}, SUS scores were computed, yielding an average of 87.08 (SD = 5.48). This significantly surpasses the average SUS score of 68, indicating excellent usability. Participants found our plug-in easy to use, with design features enhancing their GUI design process.

%The SUS scores were computed according to established practices~\cite{brooke1996sus}. The average SUS score is 87.08 (SD = 5.48), which is significantly higher than the average SUS score 68, indicating good usability. The results indicate that participants generally found our plug-in tool easy to use, and the design features are useful for their GUI design process. 

\subsubsection{Ratings of Plug-in Features} 

Participants rated each plug-in feature on a scale of 1 to 7, with the model scoring 6.77 for preview (SD = 0.47), 5.83 for element prediction (SD = 0.69), and 6.17 for confidence level (SD = 0.69). Participants prefer features of prediction confidence and previews before actual element placement.

%We requested participants to evaluate each of our plug-in features on a scale of 1 to 7, including the element prediction preview, the element prediction, and the confidence level for the prediction. On average, our model scored 6.77 out of 7 for the preview (SD = 0.47), 5.83 for element prediction (SD = 0.69), and 6.17 for confidence level (SD = 0.69). This is evidence that designers prefer features that can help them identify the confidence in the prediction and that give them a preview before actual element placement.

\subsubsection{Ratings of Result GUIs} 

Participants provided an integer score out of 7 for each GUI design, with no significant differences between conditions in ratings for completed GUIs ($t = 0.396, p = 0.695$). The average rating with the plug-in was $6.28$ (SD = 0.65), and without the plug-in was $6.17$ (SD = 0.96). Participants reported having terminated their design process upon achieving satisfaction with the GUIs.
%We did not observe significant differences between the two conditions in ratings for the completed-GUI results ($t = 0.396, p = 0.695$). The average rating is $6.28$ (SD = 0.65) with the plug-in and $6.17$ (SD = 0.96) without the plug-in. Participants reported having terminated their design process only when they had achieved satisfaction with the GUIs. 

\subsubsection{Timing} 

Without our plug-in, the average finishing time was approximately 5.6 minutes (SD = 1.02), while with the plug-in, times mostly ranged from 1 to 3 minutes (Mean = 2.22, SD = 0.78). The significant difference ($t = -11.71, p < .001$) indicates a 40\% improvement in GUI design task completion time with our plug-in.

%The average finishing time for each GUI without our plug-in was about 5.6 minutes on average (SD = 1.02) while the corresponding times with the plug-in lay mostly in the 1--3-minute range (Mean = 2.22, SD = 0.78). The statistical difference is significant ($t = -11.71, p < .001$). This shows that our plug-in improved the efficiency of the GUI design process.

\begin{ADD}
\def\arraystretch{1}% 
\begin{table*}[t!]
\scalebox{1}{\setlength\tabcolsep{5pt}
\begin{tabular}{lccccccccc}
\hline
\bf \add{Model} & \add{\textbf{Overall}} & \bf \add{Profile} & \bf \add{Menu} &\bf  \add{Login} & \bf  \add{Settings} & \bf  \add{Tutorial} & \bf \add{Form} & \bf \add{Gallery} & \bf \add{List}  \\ \hline
\add{ResNet50}  & 28.35 & 0.00 & 16.10  &  0.00   &  81.65  & 0.00 & 0.00 & 0.00 & 47.62   \\
\add{Nearest Neighbors}  & 30.62 & 65.96  &  46.19   &   43.33 & 14.75 & 65.57 & 68.00 & 19.01 & 7.44   \\ 
\add{Random Forest}  & 82.39 & 58.51  &  95.34  &   70.00 & 80.22 & 54.10 & 69.33 & 86.97 & 91.07   \\ 
\hline
\bf\add{Ours}  & \bf91.53 & \bf79.79  & \bf 96.19   &  \bf 75.00 &\bf 85.97 & \bf90.16 &\bf 88.67 &\bf 97.89 & \bf 95.24 \\ 
\bottomrule
\end{tabular}
}
\caption{\add{GUI topic classification results: Comparing the accuracy rates of our graph-based GUI representation method and the ResNet50, Nearest Neighbors, and Random Forest models for GUI topics.}}
\Description{The results of the accuracy rates of our graph-based GUI representation method and the ResNet50, Nearest Neighbors, and Random Forest models for eight GUI topics.}
\label{tbl:classification}
\end{table*}
\end{ADD}

\subsubsection{Summary} 

The high SUS score suggests ease of use and benefits for participants in their GUI design process. Our plug-in significantly improved efficiency, reducing task completion time by approximately 40\%. Participants favored features such as element prediction preview, confidence level indicator, and element prediction. However, no significant difference in GUI quality was observed between plug-in usage and non-usage, as indicated by participant ratings of result GUIs.

%The high SUS score suggests that participants found the plug-in tool easy to use and beneficial for their GUI design process. 
%The timing comparison shows that our plug-in increased efficiency, reducing GUI design task completion time by approximately 40\%. The participants favored features such as the element prediction preview, the confidence level indicator, and the element prediction. However, we observed no significant difference in GUI quality between usage with and without the plug-in, as indicated by participant ratings of result GUIs.

\subsection{Qualitative Findings}

Alongside quantitative analysis, we performed qualitative analysis. Overall, participants gave positive feedback, especially on providing proper suggestions for GUI element prediction and omitting many manual design operations, with P2 mentioning that \textit{``suggestions for element prediction are reasonable and have saved me some time on manual editing and aligning elements.''}.  

\subsubsection{Workflow}

Participants appreciated the integration of our method as a Figma plug-in. Since Figma is a popular design software, P3 concluded, \textit{``It is very useful to have this kind of integration; I do not need to spend time learning a completely new tool to use these functions, and now I can simply use the software I normally use at work.''}  P5 held the same opinion, stating, \textit{``This plug-in doesn't interrupt my design process; it's more like an add-on that helps with my design and provides inspiration. The operations are intuitive, so there is no need for us to learn how to use it specifically.''} Some participants also praised the ease of element placement and thought the plug-in makes the design process more efficient; e.g., \textit{``I often had to place the elements one by one, but now I can just click, and the plug-in directly suggests the proper placement. It is easy and less time-consuming.''} (P6). %Although Figma supports the alignment of elements well, we still need to drag and drop elements to fit the aligned line. Simply clicking definitely saved some time on that

\subsubsection{Functionality}

%When asked what functionality in the plug-in is the most useful,
Participants highlighted the usefulness of the preview window, with P5 noting its role in exploration and inspiration: \textit{``It is interesting to see element prediction previews for each element in the preview window. I could hover over each element to get intuition as to how the GUI looks after placing it without the need to actually place it and undo it if I do not like the result.
This can be used as an exploration process to help me compare different elements intuitively without additional effort.
''} and P3 emphasized that the preview window gives designers \textit{``a good way to visualize element suggestions.''} In addition, P2 mentioned that the combination of preview and confidence level makes for better exploration: \textit{```
I used the preview to compare the predictions among elements with high confidence -- or medium if no element with high confidence exists -- to decide which one I preferred to place first. 
I do not need to think much about which element I want to place since the confidence level helped me narrow down the options.
''} Additionally, all the participants appreciated the convenience of automatically predicting alignments and sizes; for instance, P4 stated, \textit{``It works particularly well when the to-be-placed element needs to align with some existing elements or have the same size as them.''}

\subsubsection{Limitations}

While acknowledging the advantages, participants identified some limitations. Specifically, P2, P3, and P5 criticized the method's accuracy when the unplaced element does not need to align or group with any existing GUI element. Furthermore, P3 and P4 pointed out that the method provides only one suggestion per element, limiting exploration possibilities.

%Despite the advantages of our approach, participants observed some limitations to the method. Firstly, P2, P3, and P5 criticized our method's ability to generate accurate predictions if the unplaced element does not need to align or group with any existing element in the GUI. 
%Additionally, P3 and P4 pointed out that, while our method offers suggestions for each element to be placed, it provides only one suggestion per element, thus restricting the exploration possibilities.
 \begin{ADD}

\section{Other Applications} 

In addition to GUI autocompletion, we further explored other applications using our graph-based GUI representation.

\subsection{GUI Topic Classification}

GUI topic classification involves categorizing GUIs based on their topics and usage. For instance, ``Gallery'' GUIs exhibit a grid-like layout with images, while ``Profile'' GUIs display information related to user profiles or products. Our approach utilized GUI representation for classification, employing eight GUI topics derived from the Enrico dataset~\cite{leiva2020enrico}. We sampled 10,000 GUIs, comprising both complete and partial instances from the Enrico GUIs associated with these eight topics, with a maximum of 2,000 instances for each GUI topic. The dataset was split into 85\% for training and 15\% for testing.
The graph representation of each GUI was fed into a Graph Neural Network (GNN) to obtain the graph embedding, following the same process used in the autocompletion task (refer to Section~\ref{sec:autocompletion}). The classification process involved three fully connected layers and a softmax function, resulting in an accuracy rate of 91.53\%, higher than other baselines. A comparison with the ResNet50, Nearest Neighbors, and Random Forest models is presented in Table~\ref{tbl:classification}.

\setlength{\tabcolsep}{2pt}
\def\arraystretch{1}% 
\begin{figure*}[!]
\def\w{0.15\linewidth}
\def\ww{0.1\linewidth}
 \centering
\begin{tabular}{c c c c | c c c }
& Input GUI & Ours & Screen2Vec & Input GUI & Ours & Screen2Vec
\\
\bf \begin{turn}{90} 
\bf \ \ \ \  \ \ \ \  \ \ \ \  \ \ \ \add{Complete GUI}
\end{turn} & 
\includegraphics[width=\w]{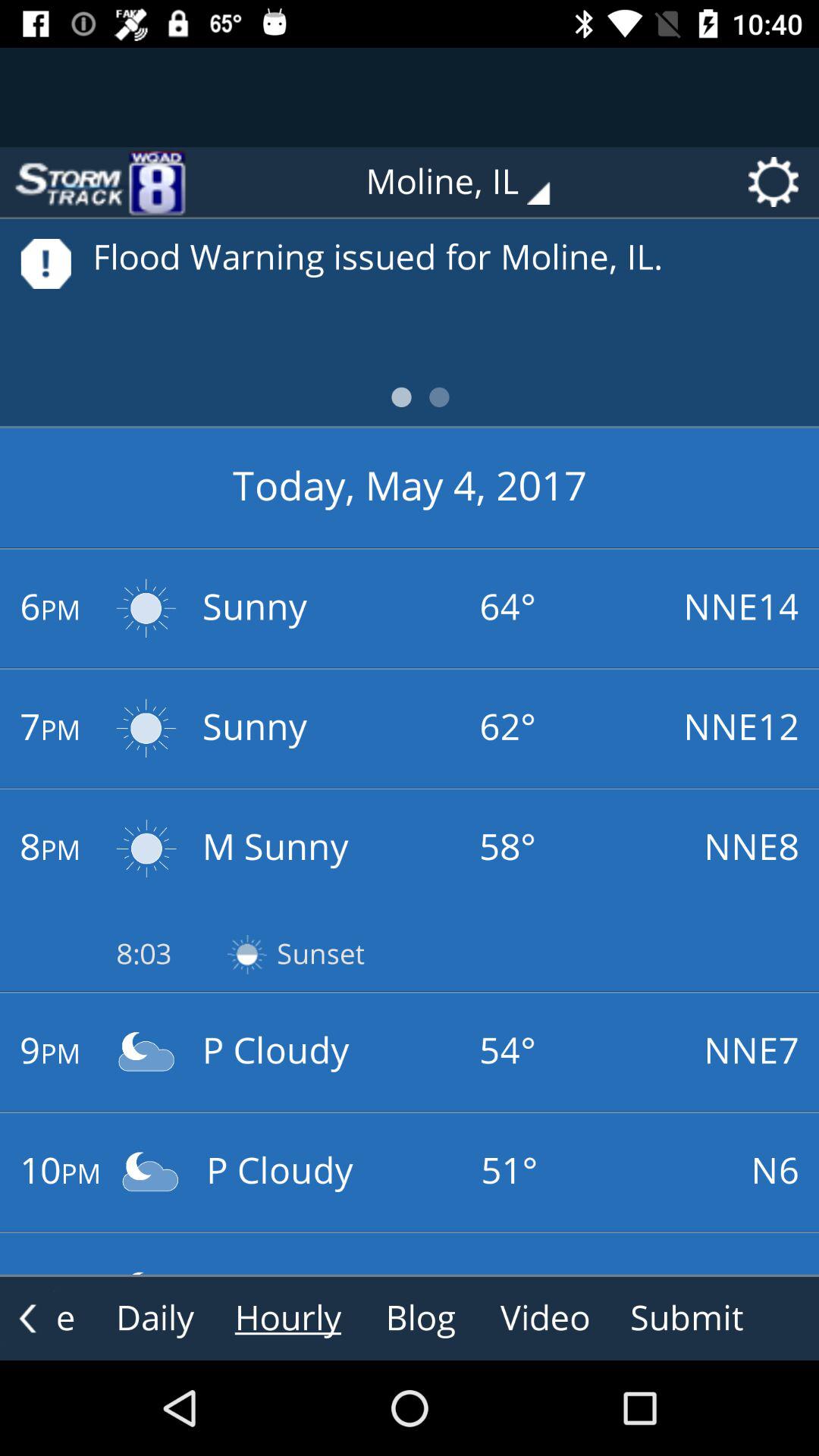} &
\includegraphics[width=\w]{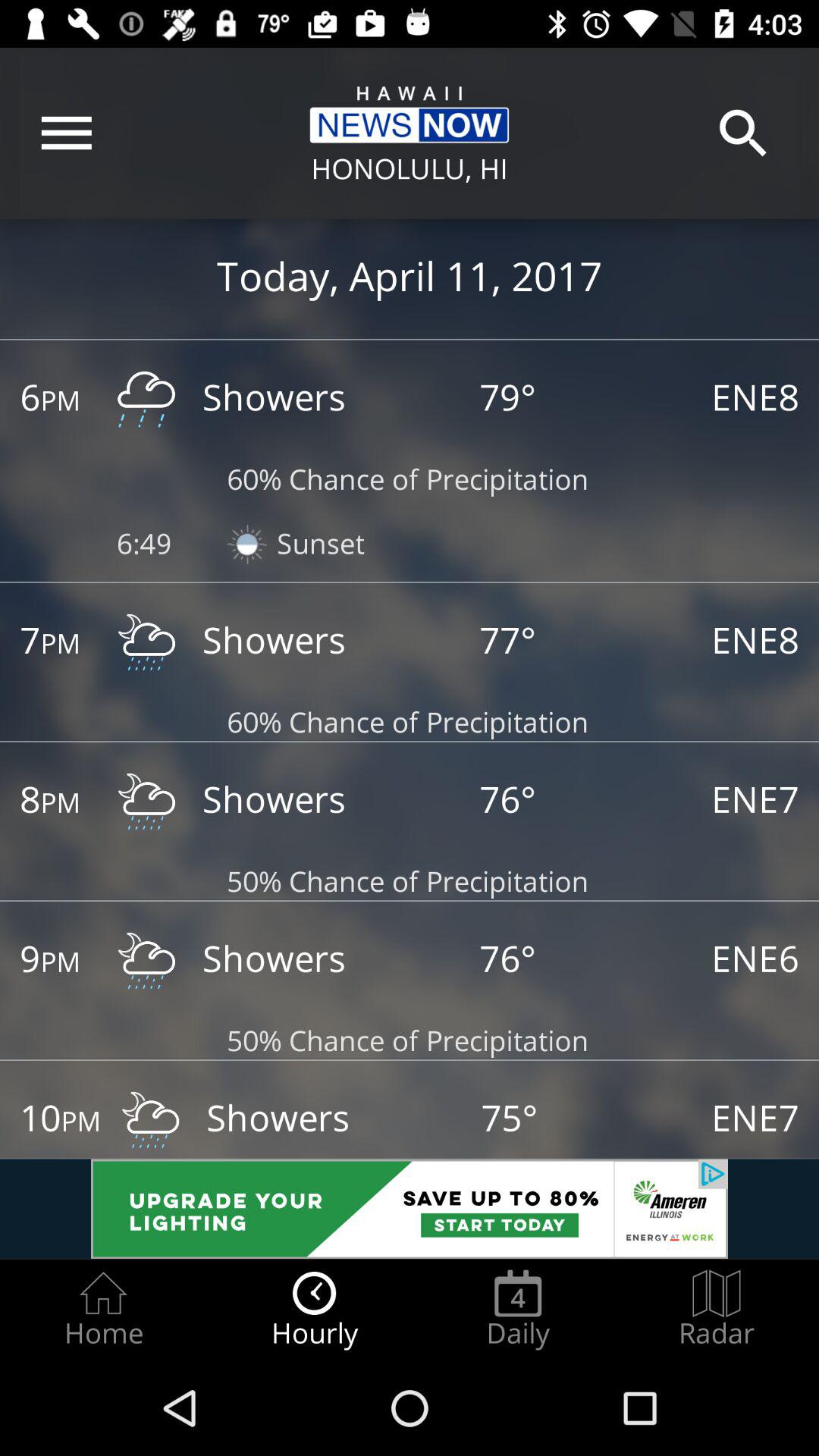} &
\includegraphics[width=\w]{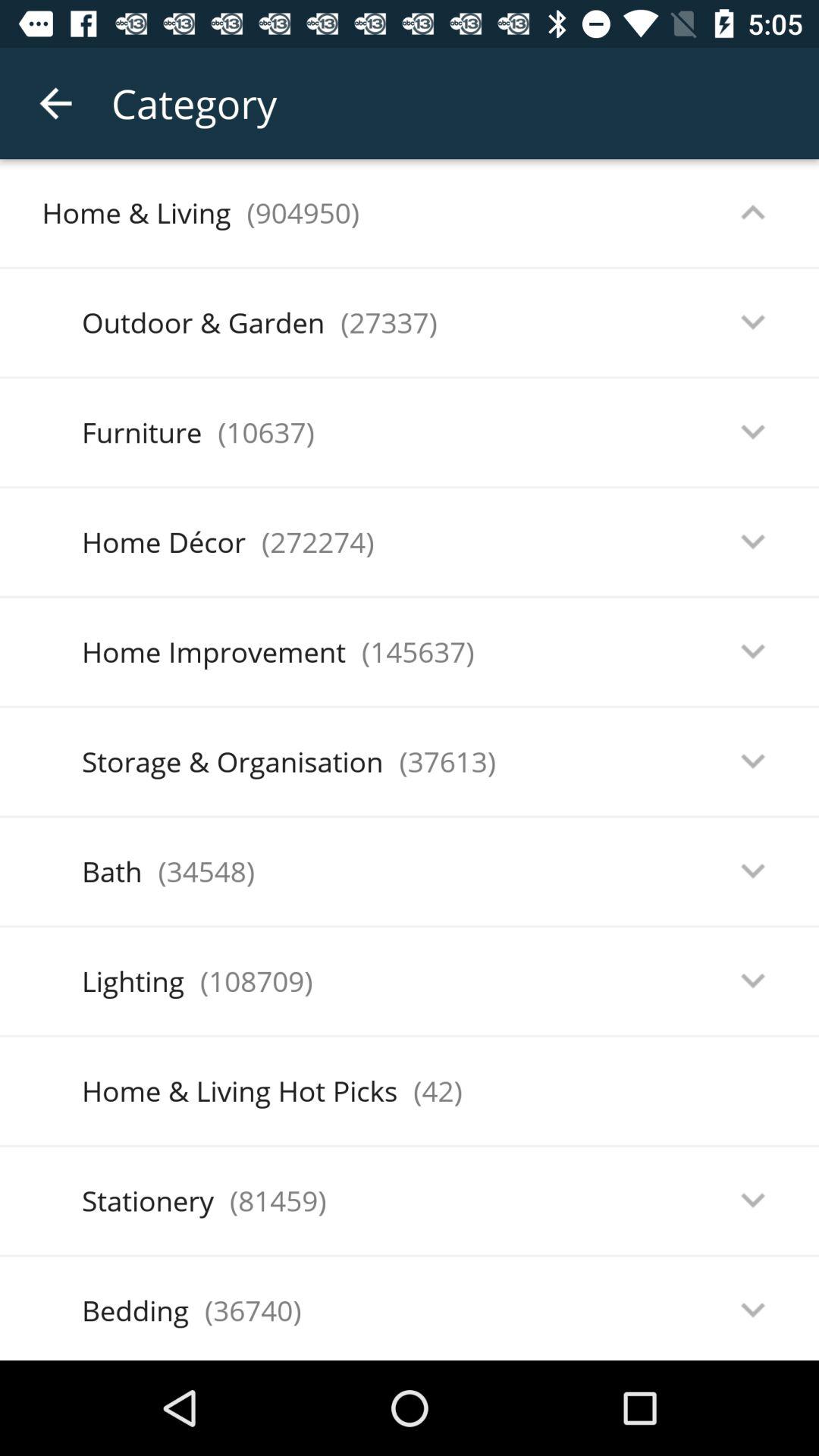} &
\includegraphics[width=\w]{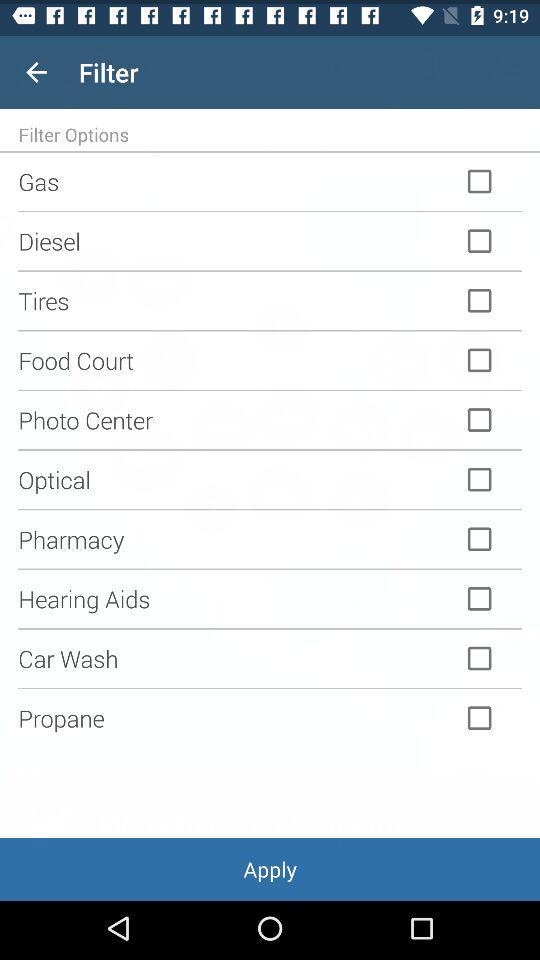} &
\includegraphics[width=\w]{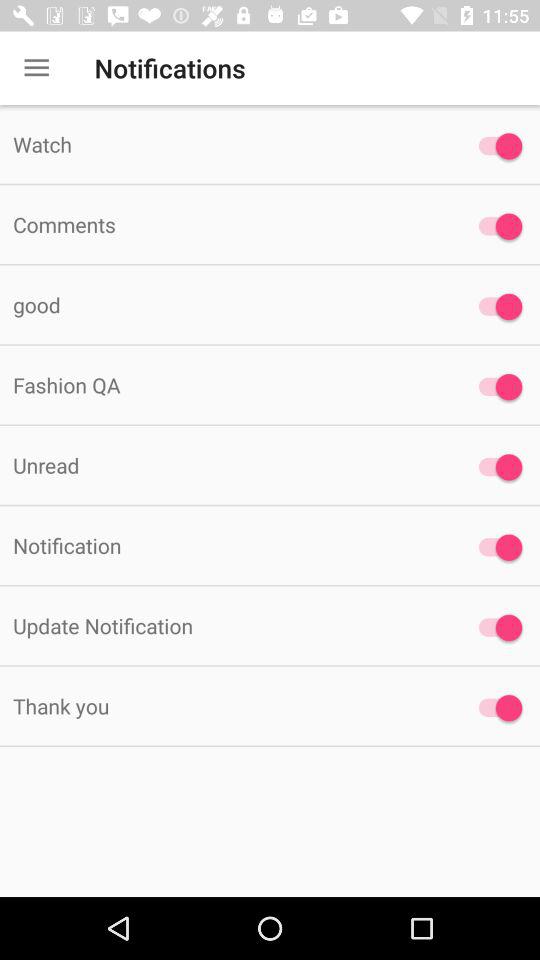} &
\includegraphics[width=\w]{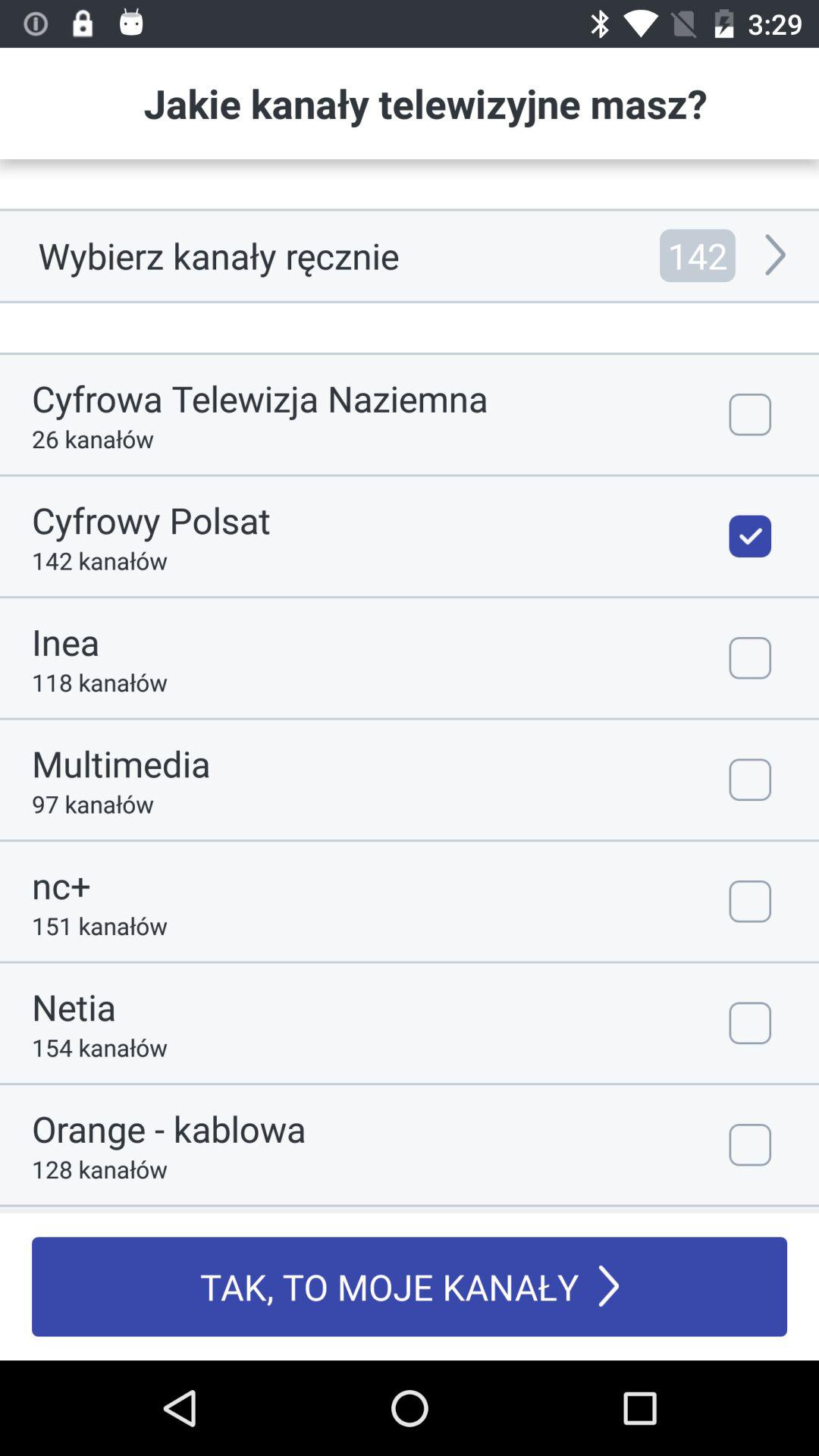} 
\\
& Input GUI & Ours & Screen2Vec & Input GUI & Ours & Screen2Vec
\\
\bf \begin{turn}{90} 
\bf \ \ \ \   \ \ \ \ \ \ \ \ \ \ \ \ \ \ \ \ \ \add{Partial GUI}
\end{turn} & 
\includegraphics[width=\w]{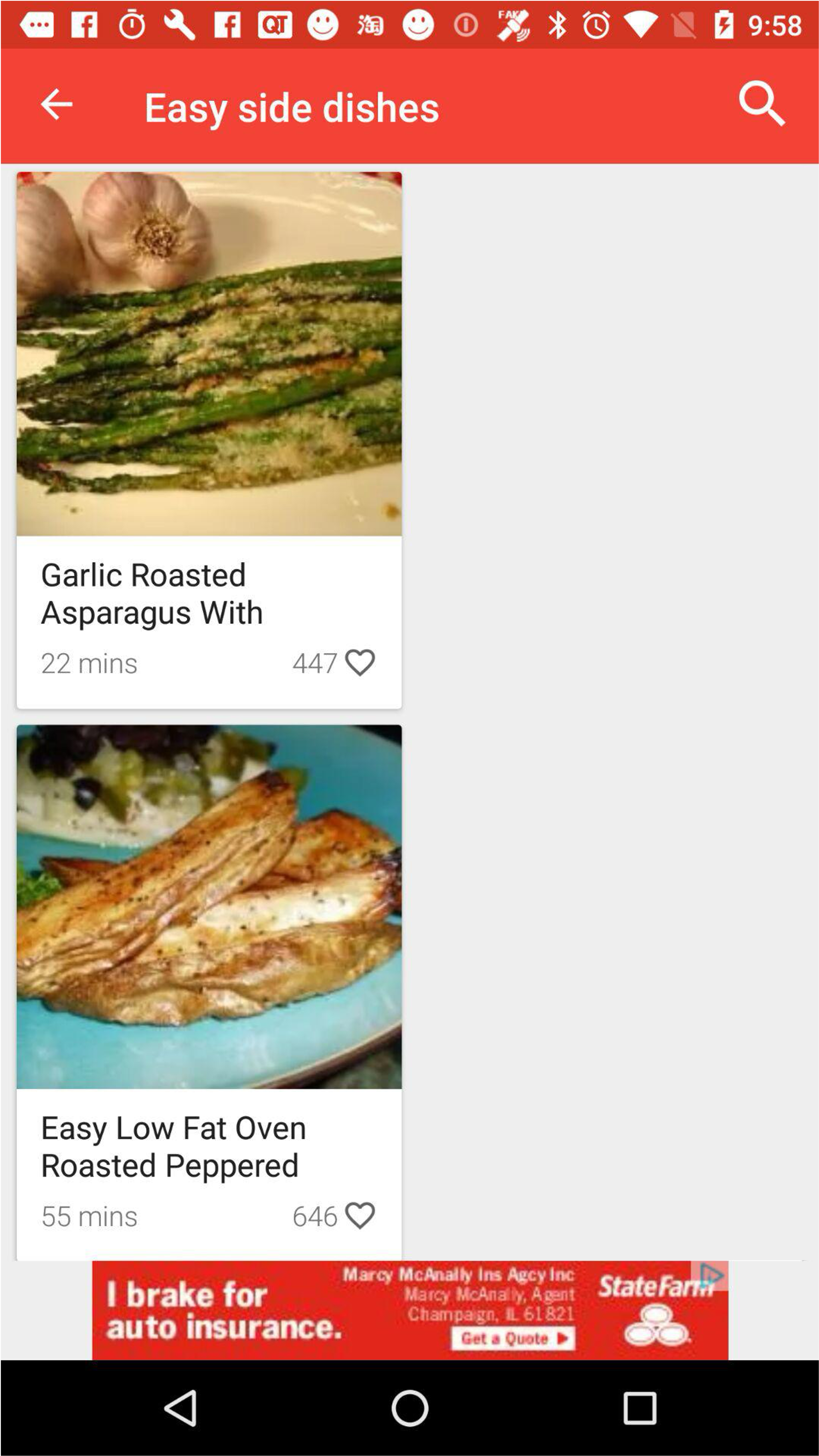} &
\includegraphics[width=\w]{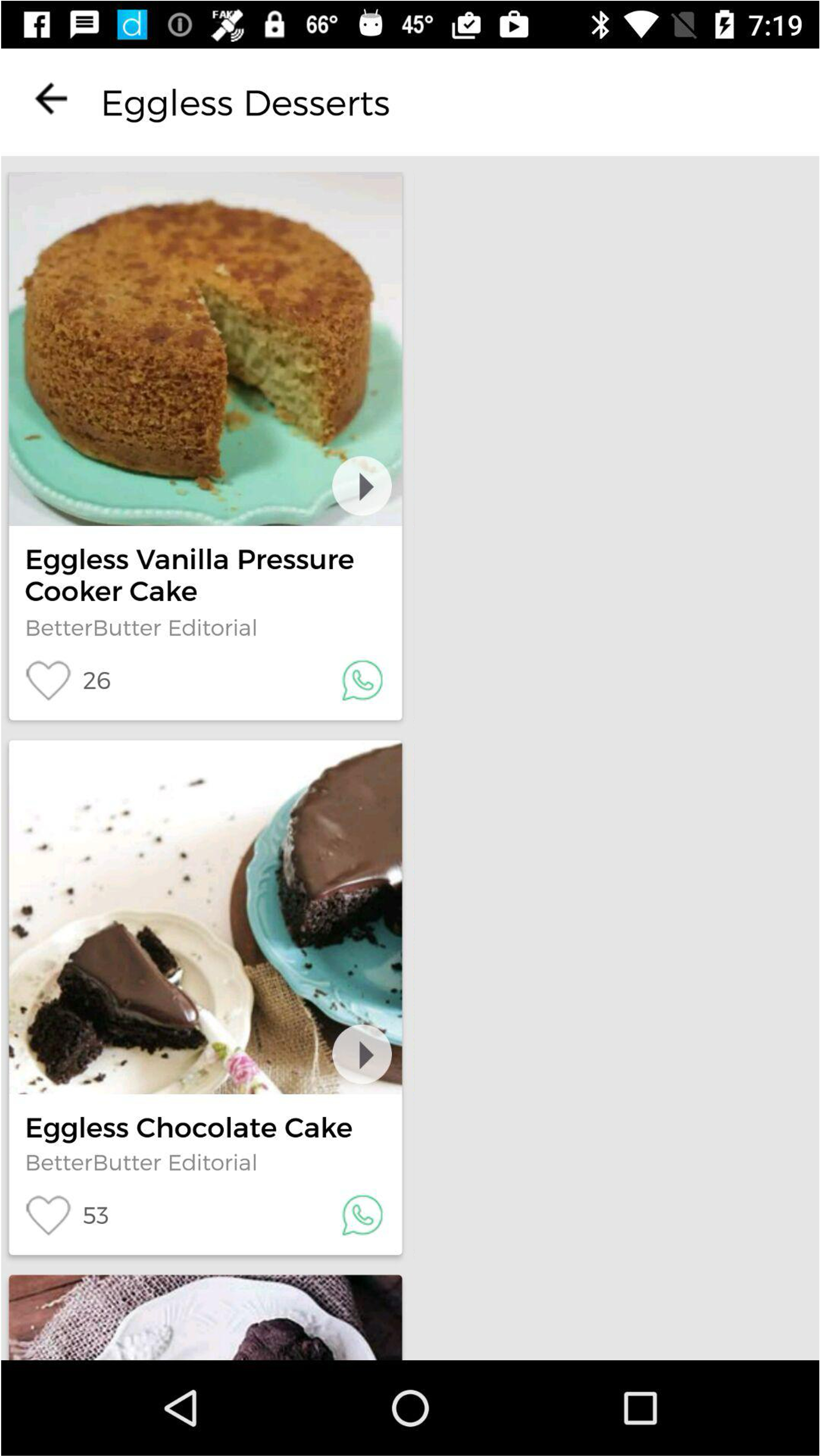} &
\includegraphics[width=\w]{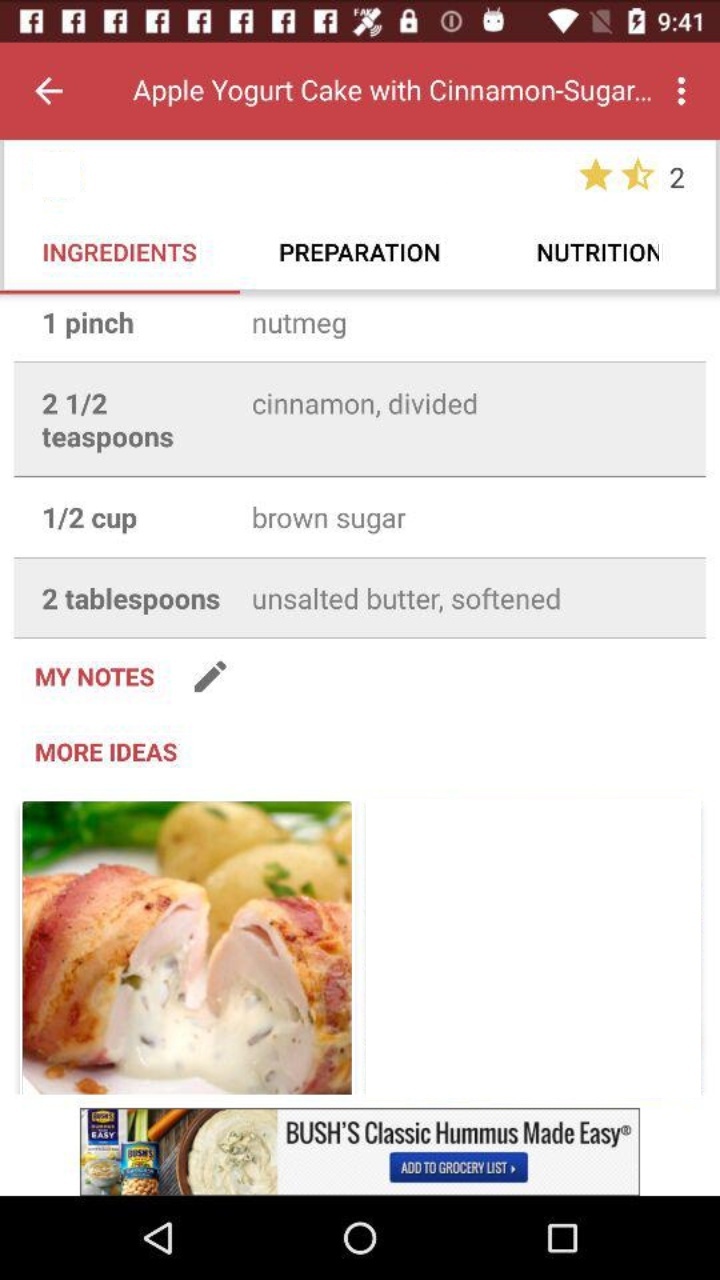} &
\includegraphics[width=\w]{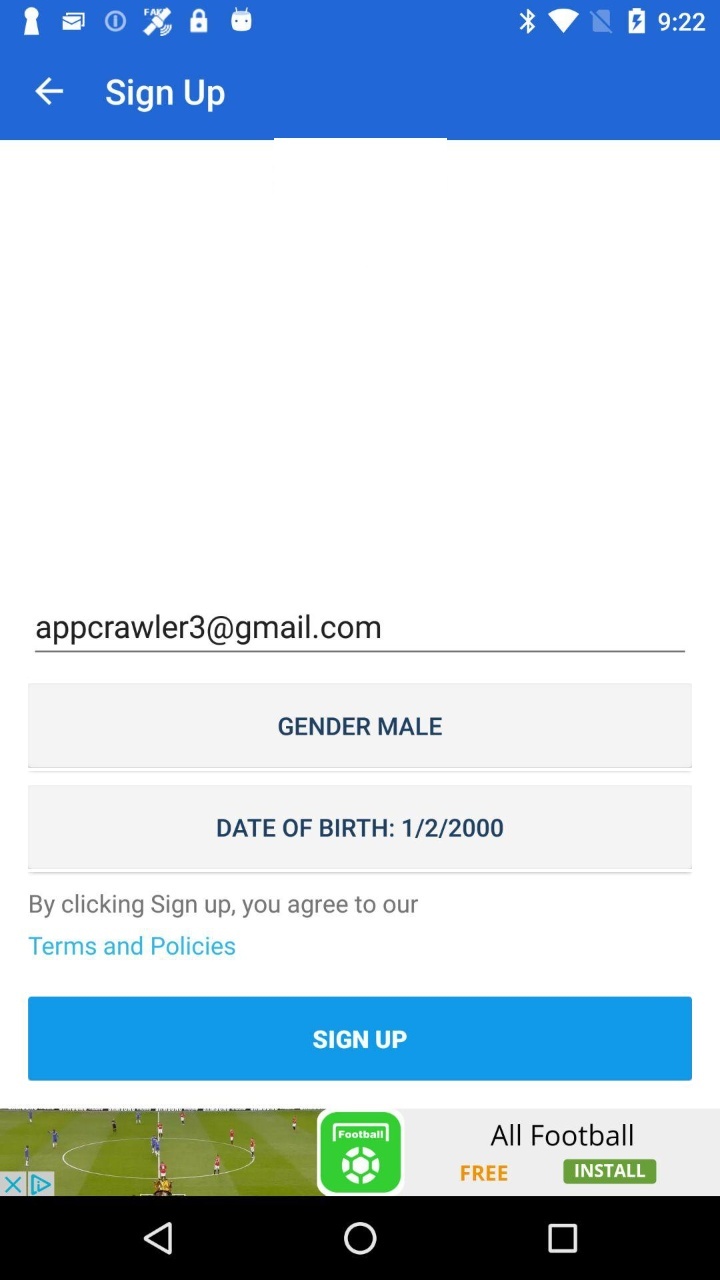} &
\includegraphics[width=\w]{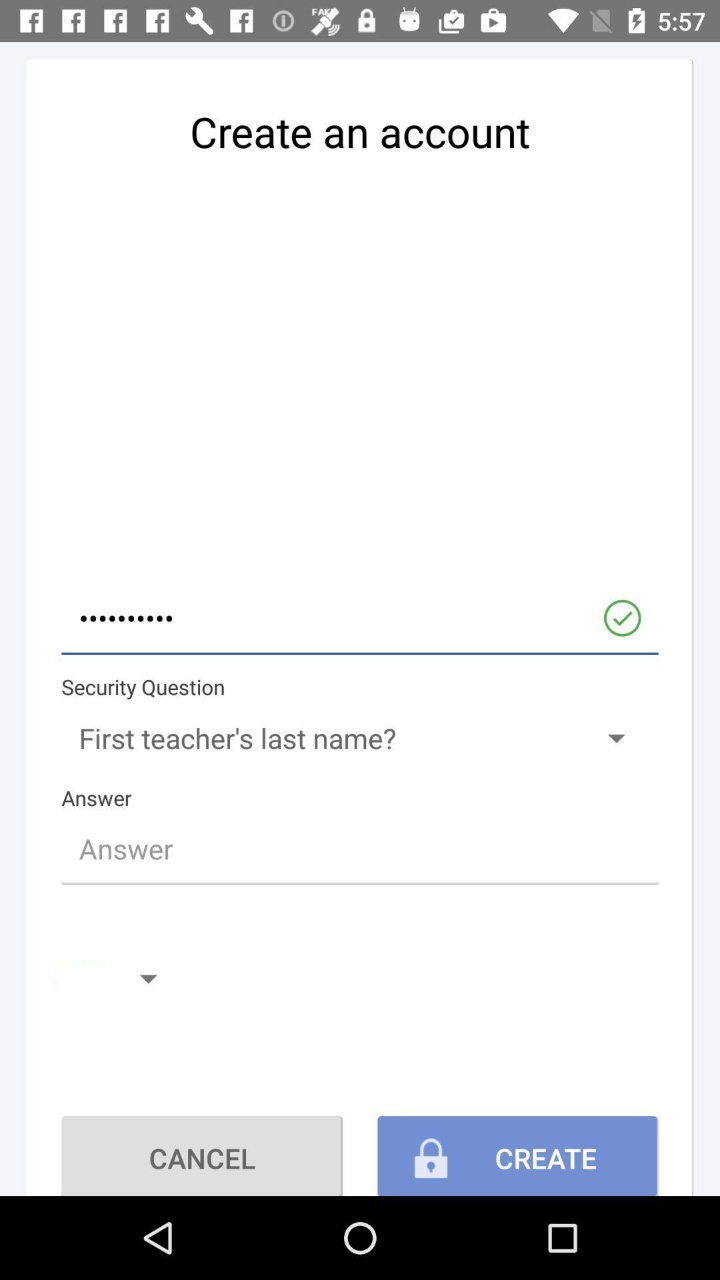} &
\includegraphics[width=\w]{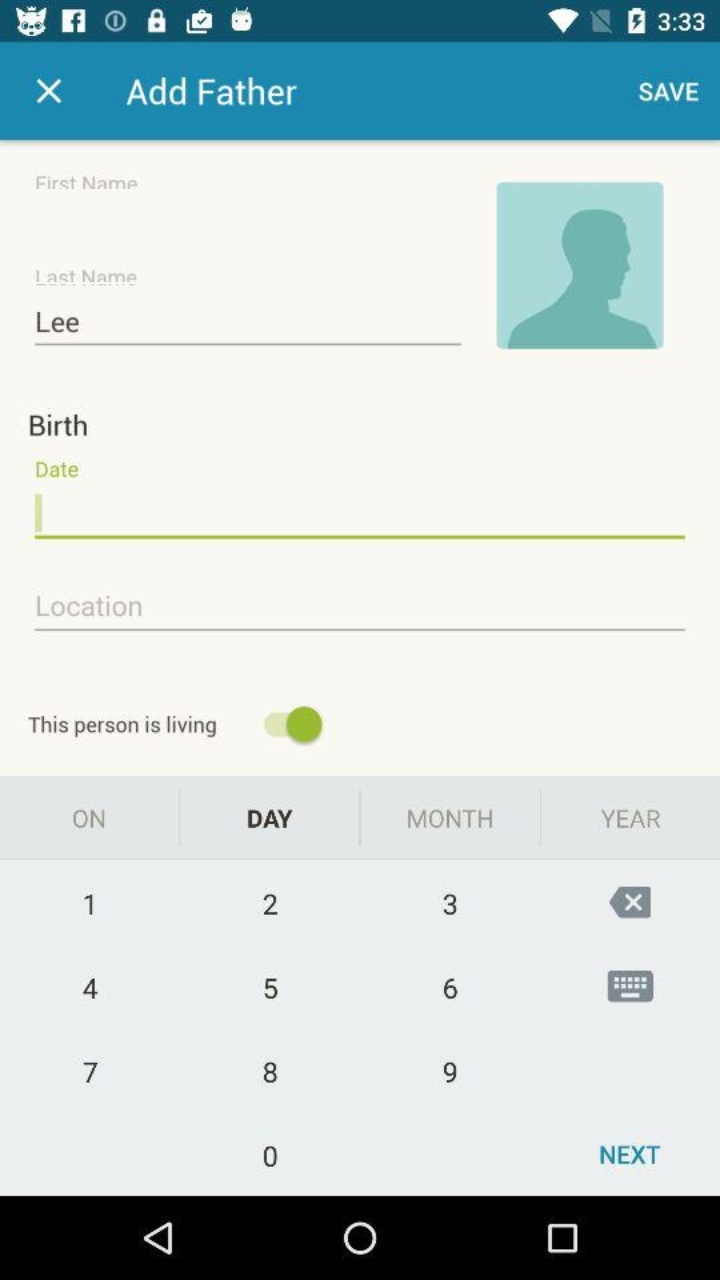} 
\\
\end{tabular}
\vspace{-10pt}
\caption{\add{GUI retrieval results of our model and Screen2Vec, including both complete GUIs and partial GUIs.}
}
\Description{GUI retrieval results of our model and Screen2Vec, including both complete GUIs and partial GUIs.}
\label{fig:retrieval}
\end{figure*}

\subsection{GUI Retrieval}

GUI retrieval is the process of finding the most similar GUI to a given one. Utilizing the graph embedding from our trained GUI topic classification model, we applied the nearest neighbor approach to identifying the closest GUIs. Samples demonstrating the performance of our model and the Screen2Vec model~\cite{screen2vec2021li} in retrieving both complete and partial GUIs are shown in Figure~\ref{fig:retrieval} (More results can be seen in the supplementary materials).

%GUI retrieval refers to the process of searching for the closest GUI to the given GUI.
%Using the graph embedding obtained from our GUI classification model described above, we employed the nearest neighbor approach to identify the most similar GUIs. \autoref{fig:retrieval} illustrates some samples of our model and the Screen2Vec model~\cite{screen2vec2021li} in retrieving both complete and partial GUIs. 

% \begin{figure}[t]
% \def\w{\linewidth}
%  \centering
% \includegraphics[width=\w]{figures/comparison_study_UI_retrieval.png}
% \caption{\add{The results of the comparison study for GUI retrieval on user preferences regarding the retrieval outcomes of our model and Screen2Vec.}
% }
% \label{fig:comparison_study_UI_retrieval}
% \end{figure}

\subsubsection{User Study}

A comparison study was conducted to assess our model against Screen2Vec.

\paragraph{Participants}

Fifteen participants (9 female, 6 male) were recruited through social media promotion. All participants had normal vision or vision corrected to normal with glasses. None were colorblind. Local regulations do not require formal ethics review.

\paragraph{Experimental Design}
From a pool of 1,500 randomly sampled partial and complete GUIs, we randomly selected 100 GUIs for each participant and presented the results retrieved by our method and the Screen2Vec method.

\paragraph{Apparatus}

Pairs of GUI images, one predicted by our method and one by Screen2Vec, were displayed side by side on a custom webpage in randomized order.

\paragraph{Procedure}

Participants began with a demographics questionnaire, followed by evaluating GUI images and selecting their preferences based on personal assessment criteria. Each participant could assess up to 100 pairs and could stop comparisons at any point.

%The participants began by filling out a demographics questionnaire. They were then asked to view the GUI images produced and select the preferred one by applying personal assessment criteria, which could involve GUI structure, type, content, or other characteristics.
%Participants indicated their preference by choosing either the left or the right image. If they found the two equally similar to the input image, they could select ``They are equally good.'' Each participant was permitted to assess, at maximum, 100 pairs and could stop the comparisons whenever wishing to before that point. 

\paragraph{Findings}

We obtained 1,144 responses from 15 participants. Preferences were as follows: 37 for Screen2Vec (3.23\%), 426 for our model (37.24\%), and 681 for images perceived as equally good (59.53\%). The difference between our method and Screen2Vec was statistically significant ($\chi^2 = 827.5, p < .001$). This indicates that our model retrieved more visually similar GUIs compared to Screen2Vec.

\end{ADD}
 \section{Discussion and Conclusion}

This paper addressed the challenges of representing GUIs through a graph-based deep learning model. Prior deep learning-based GUI representations failed to consider the constraints for GUI elements and the visual-spatial-semantic structure of a GUI, which are important in computational design. Although many modern GUI tools use constraints to optimize GUIs, training a model to predict constraints remains a challenge. 
Our proposed novel graph-based GUI representation captures both the properties of GUI elements, such as textual content, visual appearance, and element types, and their relationships in the visual, spatial, and semantic dimensions of a GUI. It can be computed efficiently in computational design. We further trained graph neural networks (GNNs) to take the graph as input to optimize the GUI. We will release our code and data.

Our work has achieved the following results in the GUI autocompletion task. 

\begin{enumerate}
\item Our method predicts the position, size, and alignment of GUI elements more accurately. As shown in Figure~\ref{fig:comparison}, it achieves less than half of the error values in these three metrics (position, size, and alignment) compared to GRIDS~\cite{dayama2020grids}, an approach for autocompletion using integer programming. When the number of existing elements on the GUI increases, it remains to have low error rates while GRIDS's errors dramatically increase. 
\item Our model offers superior alignment and visual appeal compared to the baseline method, and is better aligned with participants' preferences. In our comparison study,  70.33\% of the responses preferred results from our model compared to 13.54\% for GRIDS. 
\item Our method enhances flexibility by integrating as a plug-in within a popular existing design tool, Figma. This integration allows designers to apply workflows they are already familiar with, eliminating the need to learn new tools or switch between different design software tools. Participants in the designer study praised the plug-in for accelerating their design process without disrupting the existing functionalities of their design applications. 
\end{enumerate}

\add{In addition to the demonstrated capability of our graph-based GUI representation in the GUI autocompletion task, we show that our GUI representation can be applied to other applications, such as GUI topic classification and GUI retrieval. 
Our model demonstrated superior accuracy in GUI topic classification compared to baseline methods like ResNet50, Nearest Neighbors, and Random Forest. Furthermore, user feedback highlighted our model's effectiveness in retrieving visually similar GUIs compared to the Screen2Vec model.
Compared to other data-driven approaches, our graph-based representation facilitates the understanding of GUI structure, improving the explainability of the model. This capacity enables our representation to potentially extend to diverse downstream tasks.
For example, accessibility needs can also be represented as constraints~\cite{gajos2008improving}, and our method can train and predict layout constraints, thus it could potentially enhance accessibility. }

%beyond the autocompletion task, such as design retrieval and evaluation/improvement of accessibility. With GNNs encoding the graph representation into an embedding vector, similar GUI designs can be retrieved as they will have closer embedding vectors. 

\begin{figure}[t]
  \def\w{\linewidth}
  \centering
 \includegraphics[width=\w]{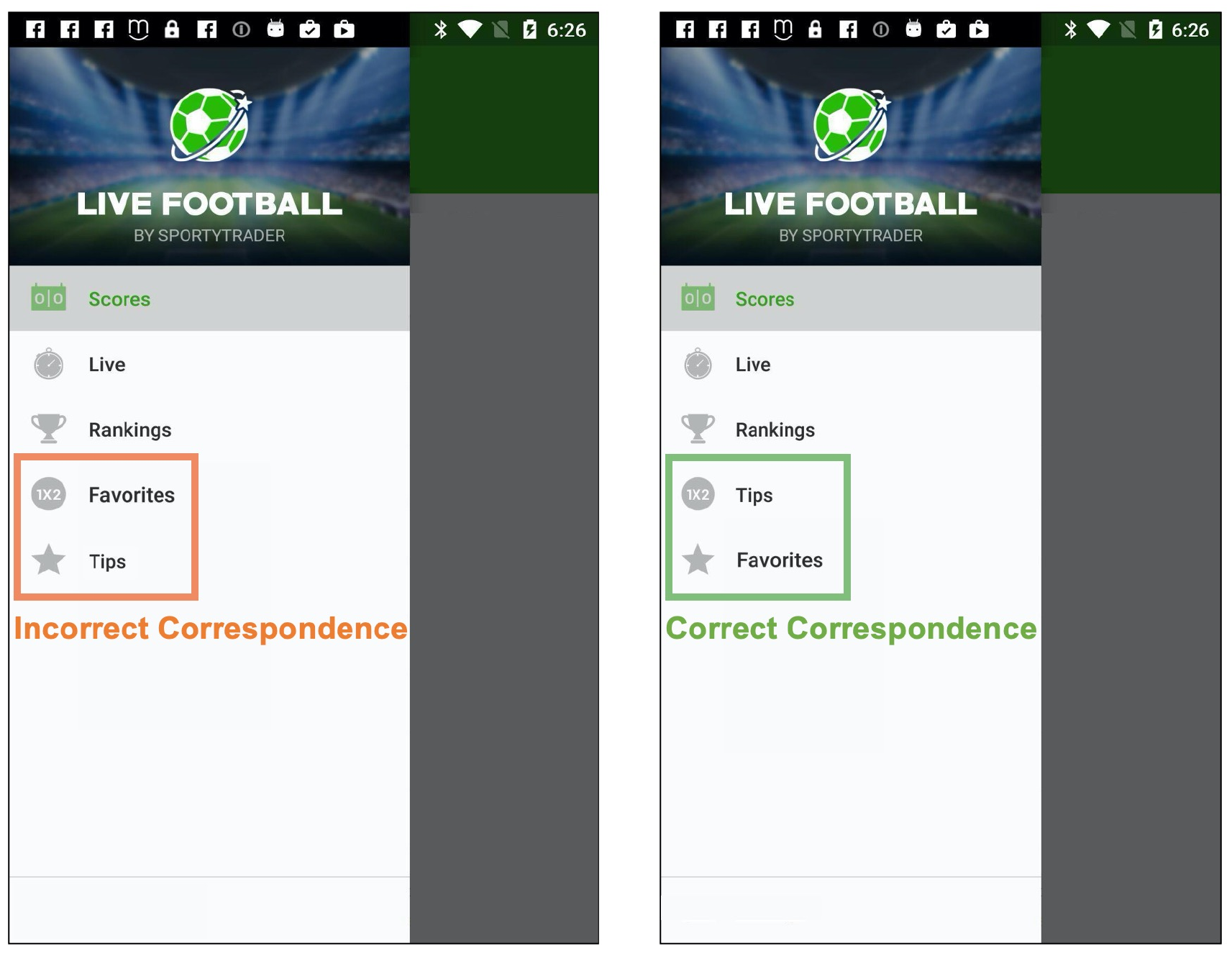}
\caption{Limitation of our method: It cannot capture the semantic correspondence between different types of GUI elements, like associating the ``Favorite'' text with a ``star'' icon, which could be explored further in future research.
}
\Description{This figure shows a limitation of our method: It cannot capture the semantic correspondence between different types of GUI elements.}
  \label{fig:failure_case}
\end{figure}

\subsection{Limitations and Future Work}

% \marginparsep=30pt
% \marginnote{\color{violet}
% R2: We clarified that we currently lack datasets containing non-rectangular bounding boxes and presented our solution for handling such boxes once they become available.\\~\\}
As pointed out by participants in our designer study, our method has limited ability to generate accurate predictions if the unplaced element does not need to align or group with any existing element on the GUI. We currently assign a low confidence level to it to avoid uncertain predictions. Future work can improve the prediction of underconstrained GUI elements by considering more design priors or including more complicated constraints.
\add{As shown in \autoref{tbl:related_work_comparison}, our representation does not explicitly represent the view hierarchy. The view hierarchy provides structural data, aiding models in understanding the layout and relationships of elements. We do not currently represent view hierarchies since they are not always available and often contain errors with incorrect structure information. However, future work can connect related element nodes in the graph representation to represent the view hierarchy.} 
Moreover, while our method offers suggestions for each element to be placed, it provides only a single suggestion per element, thus constraining the possibility of exploration.
In addition, we focus on a setting where all the elements are rectangular in shape or in rectangular bounding boxes. \add{There are no datasets available with non-rectangular bounding boxes. To accommodate various shapes of bounding boxes, we can augment the element node with additional parameters. These parameters would facilitate the description of common shapes, such as rectangles with rounded corners and circles. Subsequently, the model can be retrained to incorporate this information when present in the training dataset.} 
%Future work can explore other shapes of bounding boxes.
Furthermore, we observe that even for element prediction with high confidence levels, sometimes it does not predict ideal results. For example, as illustrated in Figure~\ref{fig:failure_case}, our method cannot capture the semantic correspondence between different types of GUI elements, e.g., it cannot detect that the ``Favorite'' text should correspond to the ``star'' icon. Future research could explore more about GUI element correspondence and constraints across UI types.

\begin{acks}

We appreciate the active discussion with Zixin Guo and Haishan Wang.
This work was supported by Aalto University's Department of Information and Communications Engineering, the Research Council of Finland (flagship program: Finnish Center for Artificial Intelligence, FCAI, grants 328400, 345604, 341763; Subjective Functions, grant 357578), and the Meta PhD Fellowship.
\end{acks}

\bibliographystyle{ACM-Reference-Format}
\bibliography{Reference}

\end{document}